\documentclass[a4paper]{article}

\usepackage[utf8]{inputenc} %unicode support
\usepackage{url}  % Formatting web addresses  
\usepackage{ifthen}  % Conditional 
\usepackage{multicol}   %Columns
\usepackage[lined,boxed,commentsnumbered]{algorithm2e}%opening
\usepackage{amssymb,amsthm,amsfonts,amsmath,enumerate,latexsym,fancyvrb} 
\renewcommand{\leq}{\leqslant}

\usepackage{tikz}

\usetikzlibrary{shapes}
\usetikzlibrary{snakes}
\tikzstyle{tre}=[circle,draw,minimum size=5.1mm]
\tikzstyle{hyb}=[circle,draw,minimum 
size=5.1mm]

\tikzstyle{indef}=[circle,draw,minimum size=5.1mm]
\newcommand{\etq}[1]{%
\draw (1) node {$1$};
}
\newcommand{\etqsm}[1]{%
\draw (1) node {\small $1$};
}

\newcommand{\etqfn}[1]{%
\draw (1) node {\footnotesize $1$};
}

\usepackage{graphicx}
\graphicspath{{images2}}

\usepackage{color}

\newboolean{publ}

\begin{document}

\title{AligNet: Alignment of Protein-Protein Interaction Networks}
\author{R. Alberich, A. Alcal\'a, M. Llabr\'es, F. Rossell\'o and G. Valiente}
\maketitle

\begin{abstract}

\noindent One of the most difficult problems difficult problem in systems biology is to discover protein-protein interactions as well as their associated functions. The analysis and alignment of protein-protein interaction networks (PPIN), which are  the standard model to describe protein-protein interactions, has become a key ingredient to obtain functional orthologs as well as evolutionary conserved assembly pathways and protein complexes.  Several methods  have been  proposed to solve the PPIN alignment problem, aimed to match conserved subnetworks or functionally related proteins. However, the right balance between considering network topology and biological information is one of the most difficult and key points in any PPIN alignment algorithm which, unfortunately, remains unsolved. Therefore, in this work, we propose  AligNet, a new method and software tool for the pairwise global alignment of PPIN that produces biologically meaningful alignments and more efficient computations than state-of-the-art methods and tools, by achieving a good balance between structural matching and protein function conservation as well as reasonable running times.
\end{abstract}

%
%\begin{keyword}
%\kwd{protein-protein interaction network}
%\kwd{global alignment}
%\kwd{network matching}
%\kwd{functional consistency}
%\end{keyword}

\section*{Background}

The regulation of cellular processes is one of the most enigmatic topics in cell biology~\cite{kitano2002systems}. The activity of cellular life relies on the proper functioning of the extremely complex networks of interactions among numerous intracellular constituents. It is  clear that proteins are the most active participants in those processes~\cite{howell1992protein}, and this is the reason why researchers invest so much effort in the study of  proteins, from their shape classification to their interactions networks, trying to unveil their function.

The increasing amount of available data in almost all biological research areas and, in particular, on protein structure and protein-protein interaction networks (PPIN), encourages data analysis as an important tool to derive biological meaning. To understand the mechanisms of protein-protein recognition at the molecular level and to unravel the global picture of their interactions in the cell, many experimental techniques have been developed so far. Some methods characterize individual protein interactions, while others screen for interactions at a genome-wide scale. 
%The most popular experimental techniques to identify protein interactions are the yeast two-hybrid method, mass spectroscopy, and gene co-expression, as well as computational methods.
Some good recent surveys on this topic are~\cite{PPI-review,Proteomic-review,keskin2016}.
%\cite~{Shoe:PanchI,Shoe:PanchII}.  These technologies have generated large 
%protein interaction networks for several model organisms, such as 
%%verb actualitzar  H. pylori\cite{Rain-ea}, S. cerevisiae\cite{uetz.ea:2000}, 
%C. elegans\cite{Li2004}, and D. melanogaster\cite{Giot2003}.

In order to analyze and contrast the available data on PPIN, several pairwise  alignment algorithms have been defined in the last 15 years. The early alignment algorithms detected similarities between small subnetworks~\cite{Kelley2004,Koyuturk2006,Li2006,Liang2006,Narayanan2007}.
PathBLAST~\cite{Kelley2004} is a tool developed to search for specific pathways 
in a PPIN. In contrast to PathBLAST, 
NetAlign~\cite{Liang2006} is a web-based tool designed to identify the conserved 
network substructures between two PPIN. The method used 
in this tool and in the algorithm presented in~\cite{Narayanan2007} is the matching of isomorphic subgraphs. The pairwise alignment method introduced 
in~\cite{Koyuturk2006}, MaWISh, produces a local alignment of two PPIN by evaluating the similarity of their graph structures 
through a scoring function that accounts for evolutionary events, while the algorithm introduced
in~\cite{Li2006} to align PPIN is based on both protein sequence similarity and network topology 
similarity, and it uses integer quadratic programming. In addition, a new and 
efficient approach to obtain multiple local alignments is described in~\cite{Flannick2006}, based on conserved functional modules. All these methods and algorithms are able to detect and obtain similarities between subnetworks.  Actually, the aim of local alignment algorithms is to find  regions with the same network structure in the  networks under comparison. For every region in one network, an alignment  with some region in the other network may be obtained, but it may happen that these local alignments are mutually inconsistent, because the same protein in one network may be matched by different local alignments to different proteins in the other network: then, as a final result, it may happen that these local alignments cannot be extended to a global alignment of the pair of PPIN.

In contrast to these local alignment methods, a global alignment algorithm is aimed at finding the best overall alignment between whole PPIN \cite{elmsallati2016global}. A global alignment is then a matching mapping between the sets of proteins of two PPIN, in such a way that each protein in one network is matched to one, and only one, protein in the other network.  
The motivation to perform a global network alignment is to compare interactomes, and to understand cross-species variations  \cite{guzzi2017survey}. Global network alignment is also related to the detection of functional orthologs \cite{Singh2008}. The identification of orthologous groups is useful for genome annotation, studies on gene/protein evolution, comparative genomics, and the identification 
of taxonomically restricted sequences. Nevertheless, there are often proteins in a network that have no biologically meaningful correspondence in another network and, thus, a meaningful alignment between a pair of networks should not necessarily cover all of them.

The first algorithm for the global alignment of PPIN was IsoRANK~\cite{Singh2008}. This algorithm produces a matching between a pair of input networks, based on the idea that a protein in one network should be matched to a protein in the other network if, and only if, the neighbors of the two 
proteins can also be matched. In order to obtain the matching, the algorithm associates a score with each possible pair of nodes of the two networks, capturing the similarity of their neighborhoods. Then, the highest scoring matching is obtained. Thus, the final result of the IsoRANK algorithm is a global alignment of two networks, but the idea behind this algorithm is that two networks are similar if the network topologies are similar. However, since nodes in these networks correspond to proteins,  the protein similarity should also be taken into account when matching two proteins; for instance, if two proteins share a similar sequence and have a similar topology in the corresponding networks, then their matching probability should be higher than for those with very different sequences.

The right balance between network topology and biological information is one of the most difficult and key points in any PPIN alignment algorithm. In fact, IsoRANK considers the possibility of taking biological information of the nodes (proteins) into account, by tuning a parameter in the score values. Several other algorithms for the global alignment of PPIN have been proposed based on the idea that ``two nodes are similar if their corresponding neighbors are so,'' and hence considering mainly  network topology but also some biological features~\cite{Liang2006,SPINAL,Neyshabur2013,Patro2012,Hashemifar2014}. As a result, some of them obtain  a high number of conserved interactions but a very low functional consistence between the matched proteins \cite{Neyshabur2013,Patro2012}. Indeed, as it is stated in~\cite{Comparison}, after performing a comparison of existing algorithms for the pairwise alignment of PPIN, the evaluated algorithms have dramatic differences in the quality of the alignments they produce, either yielding good topological or good biological matchings, but few of them do well in both aspects. In addition, they are not efficient from the computational point of view: for some of them, the software system is not well organized and they can run out of memory or spend a lot of time. Moreover, even if they do produce alignments, tend to be meaningless since the coincidences among them are very poor.

Consider  also the  analysis of eight recent aligners  (NATALIE~\cite{NATALIE}, SPINAL~\cite{SPINAL}, PISwap~\cite{PISWap}, MAGNA\cite{MAGNA}, HubAlign~\cite{Hashemifar2014}, L-GRAAL~\cite{LGRAAL}, OPTNET\cite{OPTNET}, and Module\-Align\cite{ModuleAlign}) performed in \cite{Ulign},  where several topological scores (node coverage, topological coherence, induced conserved substructure and symmetric sub-structure) and different biological coherence scores (KEGG pathway annotations, Gene Ontology annotation) are considered. The authors of  this study conclude that the agreement between the alignments produced by any two different aligners is very low (around $20\%$) and also that the topological scores are not in agreement with the biological coherence of the alignments. Even more, when the alignment process is guided by topological information only, they produce alignments with the highest topological coherence but the lowest biological coherence. In contrast, when alignments are guided by sequence information only, they produce alignments with the highest biological coherence but the lowest topological coherence. This becomes extremely inconvenient in those aligners where the user has to choose the value of a parameter in order to specify the desired balance between the topological and the sequence similarities.

Therefore, the election of the right alignment tool depends on the purpose  of the alignment itself. If the aim of the alignment is to infer biological information  about relationships between the proteins in the networks, then the aligner with highest functional coherence score must be considered. If the user is interested in finding  conserved network substructures, then the aligner with highest topological score must be considered. However, it should be noticed that when we try to obtain a global alignment between  two topologically different networks, like for instance a dense network with a high number of interactions and a sparse network with a low number of interactions, the maximum expected value of any topological score is very low. Thus, the topological score as a measure of  alignment correctness  should be considered mainly to detect small conserved subnetworks,  while, instead,  the functional coherence score should be considered as the best measure of correctness in a global network alignment,  whose goal is the detection of functional orthologs.
 
Motivated by the lack of well-balanced and efficient algorithms, we have designed AligNet, a parameter-free PPIN alignment algorithm aimed at  filling the gap between efficient topologically and biologically meaningful matchings. The overall idea of the algorithm is to obtain many local alignments that are combined and extended into a meaningful global alignment. The final alignment  captures the benefits of  considering both categories of alignments. With the local alignments we capture the topological similarity between the networks and we speed up the running time of the algorithm, while with the final global alignment we solve the inconsistencies among  the local alignments and yield an overall alignment of the pair of input PPIN. The results obtained with AligNet and with the best aligners compared in~\cite{Comparison} and in~\cite{Ulign} show that AligNet indeed achieves a
good balance between topological and biological matching. In the tests reported in this paper, AligNet obtained the highest functional coherence scores in most of the alignments, which means that it maximized the  functional consistence between aligned proteins, and also a reasonable fraction of conserved interactions. In addition, HubAlign and AligNet had   the best running times among all the aligners considered in the aforementioned tests.

\section*{Methods}

\subsection*{Protein-protein interaction networks as graphs}

A protein-protein interaction  network  (PPIN) is modelled in a  natural way as a graph, with 
its nodes representing the network's proteins and its edges, the interactions between them. Moreover, the interaction between two 
proteins is considered a symmetric property, that is, if a protein $p_1$ 
interacts with another protein $p_2$, then it is tacitly understood that   $p_2$ also interacts 
with $p_1$. Hence, PPIN are specifically  modelled by means of undirected graphs. In this way, the problem of  aligning pairs of PPIN is 
translated into the problem of  aligning pairs of undirected graphs with their nodes injectively labelled by proteins.

Formally, an (\emph{undirected}) \emph{graph} is a structure $G=(V,E)$ with $V$ a finite set of \emph{nodes}  and $E$ a family of 2-element subsets $\{u,v\}$ of $V$, called the \emph{edges} of the graph; recall that, as sets, $\{u,v\}=\{v,u\}$.
We say that an edge $e=\{u,v\}$ \emph{connects} the nodes $u$ and $v$,  and also that $e$ is \emph{incident} to $u$ and 
$v$. The nodes $v$ such that $\{u,v\}\in E$ are the \emph{neighbors} of $u$.
We shall denote by $N_G(u)$ the set of neighbors of $u$ in $G$.

We introduce now some further definitions and notation that will be used throughout this 
paper. Let $G=(V,E)$ be an undirected graph. 
\begin{itemize}
\item The  \emph{degree} of a node $u\in V$ is the number of edges 
incident to it; that is, the cardinal of $N_G(u)$. We denote it by $\deg(u)$. 

\item A \emph{path} between two nodes $u,v\in V$ is a sequence of pairwise different edges
$\{u,u_1\},\{u_1,u_2\},\allowbreak \ldots,\{u_{k-1},u_k\},\{u_k,v\}$ such that the first and last edges are incident to $u$ and $v$, respectively, and every pair of consecutive edges share a  node (different from $u$ and $v$, in the case of the  first and last edges, respectively). Two nodes are 
 \emph{connected} when there exists a path between them. The \emph{length} of a path is the number of edges forming it, and its \emph{intermediate nodes} are $u_1,\ldots,u_k$.

\item For every pair of connected  nodes $u,v\in V$, their \emph{distance} in $G$ is 
the length of a shortest path connecting them. We denote it by $d_G(u,v)$.

\item The \emph{diameter} of $G$ is the maximum distance between any two connected nodes in 
$G$. We denote it by $D(G)$.

\end{itemize}

Figure~1 displays  two  toy  PPIN that will be used as a running example throughout this section. The first network consists of $8$ nodes and $9$ edges, while the second network consists of $9$ nodes and $17$ edges.

\begin{center} 
\begin{figure}[h]    
%\label{InputNetworks1}  
\includegraphics[scale=0.8]{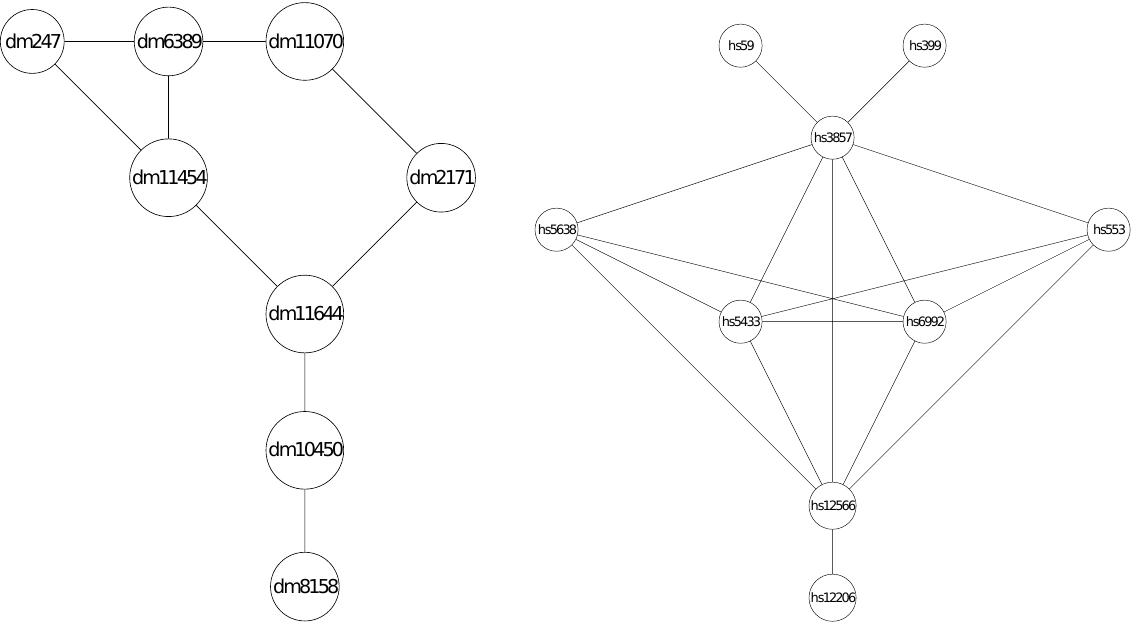}
\caption{This figure shows two small pieces of  PPINs  that we shall use to visualize the performance of AligNet. The subnetworks belong to the \emph{Drosophila melanogaster} (dme)  and the  \emph{Homo sapiens} (hsa) PPINs contained in the  IsoBase database. The first network  has  $8$ proteins  and $9$ interactions, and the second network has $9$ proteins and $17$ interactions. The diameter of the first network is $4$ and the diameter of the second network is $3$.}
%A subnetwork of the \emph{Drosophila melanogaster} PPIN (left).  A subnetwork of the \emph{Homo sapiens} PPIN (right).}
\end{figure}

\end{center}

\subsection*{The structure of the AligNet algorithm}
AligNet receives as input two graphs $G$ and $G'$ 
representing two PPIN (in particular, each node of them is labeled with a protein, in such a way that different nodes in a graph correspond to different proteins) and  produces, as output, a similarity score for them
and a local and a global alignment 
between them.  AligNet has been implemented in R \cite{CRAN}, and the implementation
is freely available from~\url{http://bioinfo.uib.es/~recerca/AligNet/}.

The main steps in AligNet are:
\begin{enumerate}
\item[1] The computation of overlapping clusterings $C(G)$ and $C(G')$, respectively, of the input networks $G$ and $G'$.
\item[2] The computation of alignments between pairs of clusters in $C(G)$ and $C(G')$.
\item[3] The computation of a matching between $C(G)$ and $C(G')$.
\item[4] The computation of a local  alignment of the input networks $G$ and $G'$.
\item[5] The extension of this  local  alignment  to a meaningful global alignment.
\end{enumerate}

Throughout this  section,  $G=(V,E)$ and $G'=(V',E')$ will denote two  graphs representing 
the input PPIN. We shall identify each node in any of these graphs with the protein it represents.

\subsubsection*{Step 1. Overlapping clusterings}

The first step in AligNet consists in computing an overlapping clustering of each input 
network. These clusterings are based on a specific 
similarity score $s(u,v)$ between pairs of proteins (nodes) $u,v$ in a PPIN, which is defined as follows:
for every pair of connected nodes $u,v$ in a graph $G$  representing a PPIN, 
$$
s(u,v)= \frac{B(u,v)+\frac{D(G)+1-d_G(u,v)}{D(G) +1}}{2}
$$
where:
\begin{itemize}
\item $D(G)$ is the diameter of the $G$ and $d_G(u,v)$ is the distance between $u$ and $v$.

\item $B(u,v)$ is the \emph{normalized bit score} of the proteins associated to the nodes $u$ and $v$, that is,  a rescaled version of their alignment score obtained with BLAST+, which is independent of the size of the 
search space~\cite{blast+}. 

\end{itemize}
If $u,v$ are not connected by a path, then $s(u,v)=0$.

The intuition behind this similarity score is 
that two proteins are similar if they have  similar sequences of nucleotides 
and they are relatively close to each other in the graph. Recall that two 
proteins interact  when there is an edge connecting them, 
that is, when their distance is $1$. Therefore,  the plausibility that two proteins 
have a related biological function increases when they are close to each other in the 
graph.  

To obtain the overlapping clustering of an input network, we define a cluster centered 
at every node of the graph as follows.  Let $\alpha$ be the third quartile of the distribution of the 
similarity score values of pairs of nodes: that is, $\alpha$ is the value for which only 25\% of the pairs of nodes $(u,v)$ are such that $s(u,v)>\alpha$. Then, for 
every node $u\in V$, the \emph{cluster}  $C_{u}$ in $G$ \emph{centered} at $u$ is
$$
C_{u}=\{ v\in V \mid s(u,v) > \alpha\}.
$$
We denote by $C(G)$ the set of clusters of a PPIN $G$.

So, the first step of AligNet computes the overlapping clusterings $C(G)$ and $C(G')$ of the input networks $G$ and $G'$. As a running example throughout this section we will consider two small networks. 
Figure~2 displays  the first PPI 
network considered as a running example as well as its overlapping clustering. This first network consists of $8$ nodes and $9$ edges, so there are $8$ clusters. Figure~2 displays  the second PPI 
network which consists of $9$ nodes and $17$ edges, and its overlapping clustering has $9$ clusters. 

 \begin{figure}[h]    
%\begin{minipage}[c]{0.46\linewidth}
 \centering
 \hspace*{-0.2in}
\includegraphics[scale=0.8]{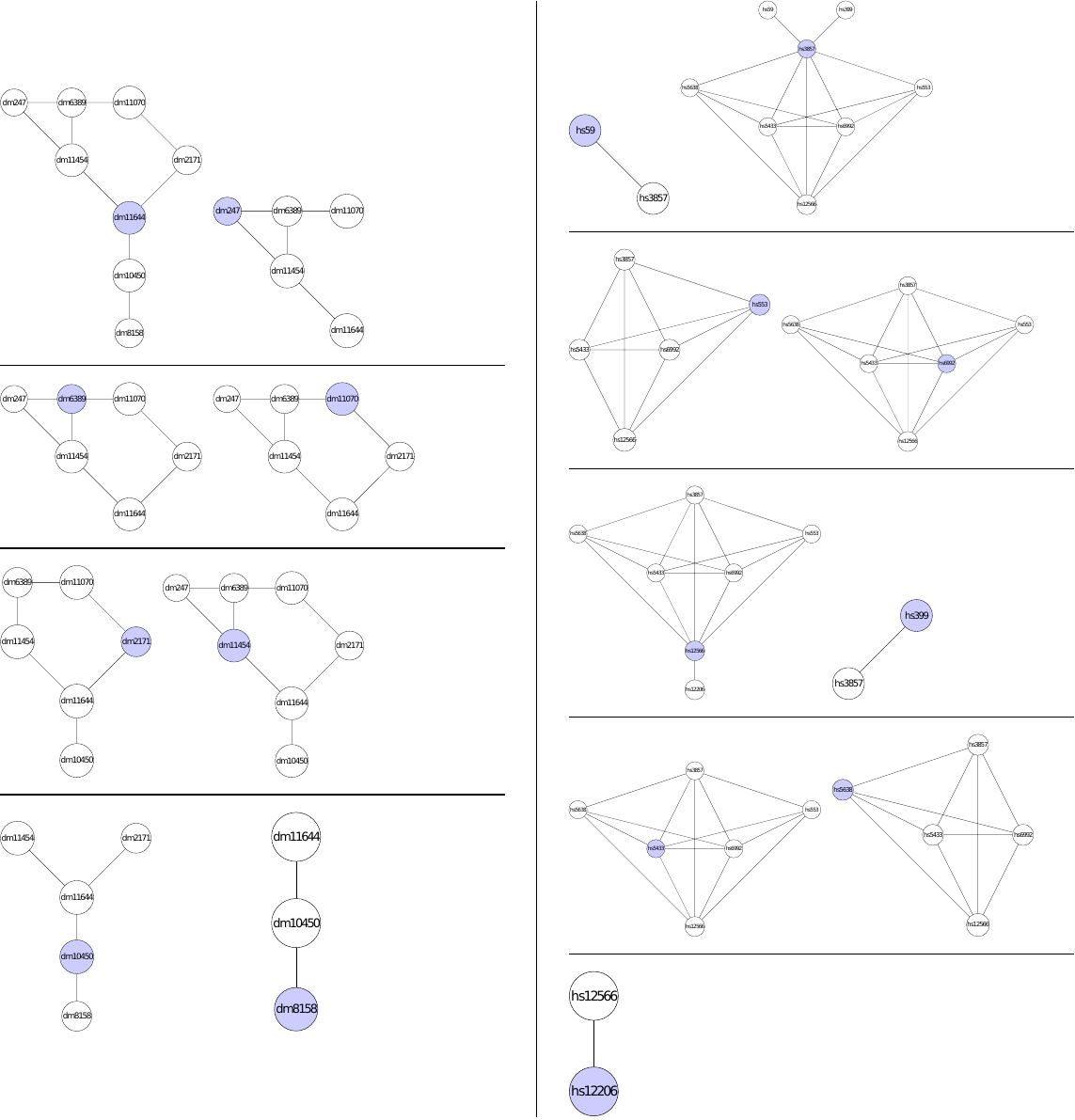}
%\end{minipage}
%\begin{minipage}[c]{0.46\linewidth}
% \centering
% \includegraphics[scale=0.35]{images2/hsaClusters.pdf}
%\end{minipage}
\caption{\label{clustering} This figure shows the overlapping clustering on the PPINs in Figure~1 obtained by AligNet.  We can 
see here  the $8$ clusters in the network in Figure~1  on the left, and 
the $9$ clusters in   the network in Figure~1 on the right.
The center of every  cluster is highlighted in blue. Since we have considered two small pieces of a PPIN, we obtain here that, the first cluster on the left is the entire piece of network. In the right, we obtain also the entire piece of network in the second cluster on the right. Notice that we obtain the whole piece of the network when we consider the cluster of a node that is in the center of the network..}
\end{figure}

\subsubsection*{Step 2. Alignments between pairs of clusters}

In this second step, AligNet computes an alignment between every pair of clusters $C_u\in C(G)$ and $C_{u'}\in C(G')$ such that $B(u,u')>0$. That is, if we assume that $B(u,u')>0$ for every pair $u\in V$ and $u'\in V'$ then AligNet computes $|V| \cdot |V'|$ alignments. These alignments define an alignment score between every such a pair of clusters that will be used in the third step to compute a matching between $C(G)$ and $C(G')$. 

The general idea to obtain the alignment between a pair of clusters $C_u\in C(G)$ and $C_{u'}\in C(G')$ (with $B(u,u')>0$) is the following: we first match the centers of the clusters, that is, we match $u$ with $u'$ and then, we match the neighbors of $u$ to the neighbors of $u'$. To decide the neighbors matching, we take into account their sequence similarity and their degrees. Thus, a neighbor of $u$ is matched to a neighbor of $u'$ provided that they have similar nucleotide sequences and also similar degrees. Following the same criteria, we match the neighbors of the neighbors of $u$ with the neighbors of the neighbors of $u'$. We iterate this process until  no unmatched neighbors are found. In the intermediate steps we keep the node matching in a list of pairs denoted by  $L_{u,u'}$. When the algorithm terminates, $L_{u,u'}$ provides a partial mapping between the nodes in  $C_u$ and the nodes in  $C_u'$.

Thus, the alignment between a pair of clusters $C_u\in C(G)$ and $C_{u'}\in C(G')$ (with $B(u,u')>0$) 
can be formally defined as follows:
\begin{itemize}
\item[(i)] Match $u$ with $u'$.  Set $L_{u,u'}=\big\{(u,u')\big\}$, 
$L_{u,u'}^{(1)}=\{u\}$ and $L_{u,u'}^{(2)}=\{u'\}$.

\item[(ii)] For every $v\in C_u\cap N_G(u)$ and for every $v'\in C_{u'}\cap N_{G'}(u')$, let
$$
F(v,v')=| \deg(v)-\deg(v')| - B(v,v') +1.
$$ 
Compute a matching $M_{u,u'}\subseteq (C_u\cap N_G(u))\times (C_{u'}\cap N_{G'}(u'))$ that minimizes $\sum_{(v,v')\in M_{u,u'}} F(v,v')$.
Sort the pairs in $M_{u,u'}$ in decreasing order of their $F$ value, 
and concatenate them to $L_{u,u'}$. Add their first coordinates to $L_{u,u'}^{(1)}$ and their second coordinates to $L_{u,u'}^{(2)}$.

\item[(iii)] Iterate step (ii), replacing $(u,u')$ by the rest of the pairs in $L_{u,u'}$ and removing from $C_u$ and $C_{u'}$ the nodes already aligned. 

More specifically, in the $k$-th iteration, take the $k$-th element $(v_0,v_0')$ of $L_{u,u'}$. For every $w\in (C_u\setminus L_{u,u'}^{(1)})\cap N_G(v_0) $ and every $w'\in (C_{u'}\setminus L_{u,u'}^{(2)})\cap N_{G'}(v_0')$,
compute $F(w,w')$. Then, compute a matching 
$$
M_{v_0,v_0'}\subseteq \big((C_u\setminus L_{u,u'}^{(1)})\cap N_G(v_0)\big)\times \big((C_{u'}\setminus L_{u,u'}^{(2)})\cap N_{G'}(v_0')\big)
$$
 that minimizes $\sum_{(v,v')\in M_{v_0,v_0'}} F(v,v')$.
 Sort the pairs forming $M_{v_0,v_0'}$ in decreasing order of their $F$ value, 
and concatenate them to $L_{u,u'}$. Add their first coordinates to $L_{u,u'}^{(1)}$ and their second coordinates to $L_{u,u'}^{(2)}$.
\end{itemize}
The matchings in step (ii) as well as in each iteration in step (iii) are computed with the Hungarian 
algorithm~\cite{Hungarian}.   Figure~3 shows an example of the alignment of a pair of clusters: one cluster from the first network and another cluster from the second network.

The overall idea behind the algorithm described above is that a node $v$ in $C_u$ should be matched to a node $v'$ in $C_{u'}$ when they  have a similar topological role in the cluster and similar sequences, provided that, furthermore, there exist paths connecting 
the cluster centers $u$ and $u'$ with $v$ and $v'$, respectively, such that their intermediate nodes are already aligned in sequential order along the paths. The alignment procedure gives priority to matching neighbors of nodes $x,x'$ at the possible shortest distance of the respective cluster centers and with $F(x,x')$ as large as possible among those pairs already matched at their same iterative step.

 \begin{figure}[h]    
 \centering
\includegraphics[scale=0.8]{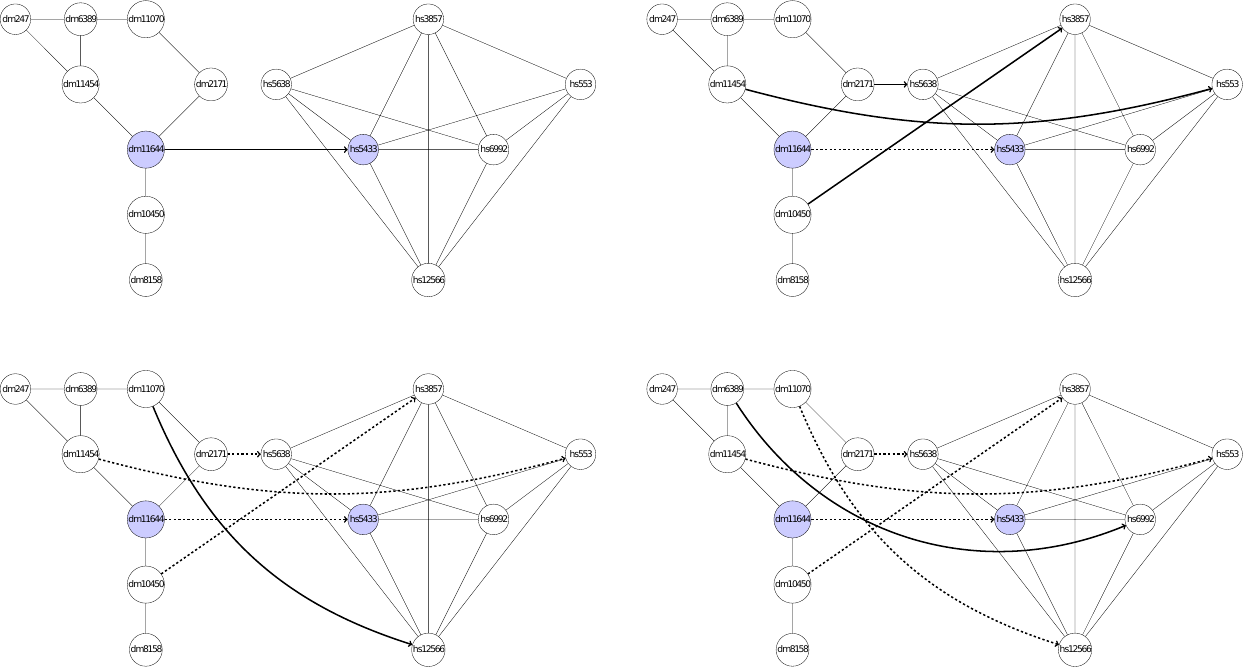}
\caption{\label{clusteralign}This figure shows how AligNet aligns  two clusters which corresponds to Step 2 of our algorithm. The clusters  in this example are, respectively, the first in the list of clusters of $G$, which are shown on the left in Figure~2 and the seventh  in the list of clusters of $G'$, which are shown on the right  in Figure~2.  We show in the picture all the steps needed to align the cluster of $G$ with the cluster of $G'$. From top to bottom in this figure, we can see that  AligNet first aligns the centers of the clusters, which are the nodes highlighted in blue.  Then,  AligNet aligns the neighbors of the centers (second row). Next, AligNet aligns the neighbors of the neighbors. In each step we show in dashed lines the nodes that are already aligned and in solid lines the nodes that are aligned in the present step. Notice that, in this example, there are two nodes that remain unmatched.}
\end{figure}

 The resulting alignment $L_{u,u'}$ 
defines a partial injective mapping  $\eta_{u,u'}: C_u\rightarrow C_{u'}$. The nodes in $C_u$ that are matched to nodes in  $ C_{u'}$ form the domain of the mapping  $\eta_{u,u'}$, which is denoted by $Dom\, \eta_{u,u'}$.

%\red{\bf The description of the algorithm is given in   Algorithm~\ref{locals}.}

\subsubsection*{Step 3. Matching between families of clusters}

Let 
$$\mathcal{A}=\{\eta_{u,u'} \mid u \in V,\ u'\in V', B(u,u')>0\}$$ be the set of  alignments obtained in step 2.
The score of every 
alignment $\eta_{u,u'}\in \mathcal{A}$ is defined  as
$$Score(\eta_{u,u'})=\frac{\sum_{v\in Dom \, \eta_{u,u'}}B(v,\eta_{u,u'}(v))}{|Dom \,\eta_{u,u'}|}+ 
\frac{|Dom \, \eta_{u,u'}|}{max_{\eta_{w,w'}\in \mathcal{A}}|Dom \,\eta_{w,w'}|}$$ 
where $|X|$ stands for the number of elements in the set $X$. This score assesses simultaneously the average similarity of the sequences of the proteins matched by $\eta_{u,u'}$ and their number.

%In an analogous way is defined the score of the alignment $\eta_u': 
%C_{u'}\rightarrow C_{u}$,   $Score(\eta_{u'})$.

Once computed all these scores,  AligNet obtains a matching between $C(G)$ and  
$C(G')$ by considering a bipartite graph where the nodes are the clusters in $C(G)$ 
 and $C(G')$, the edges correspond to alignments
$\eta_{u,u'}\in \mathcal{A}$, and the weight of an edge connecting  $C_u$ with $C_{u'}$ is the 
 corresponding score $Score(\eta_{u,u'})$. 
The matching between the nodes in $C(G)$  and  the nodes in 
$C(G')$ is then obtained by  applying the maximum weighted bipartite matching algorithm 
 to this bipartite graph.  Recall that the nodes in  $C(G)$  are the clusters in $G$ and the nodes in $C(G')$ are the clusters in $G'$. 
 The solution to the maximum weighted bipartite matching problem provides us with a matching between the clusters in $G$ and the clusters in $G'$.
 We shall denote by $\mathcal{C}$ the set of partial injective mappings $\eta_{u,u'}$ corresponding to pairs of clusters $(C_u,C_{u'})$ that are matched by this matching.

Figure~4 shows the matching between the family of clusters in 
Figure~1 and the  family of clusters in Figure~2 obtained in this step.

 \begin{figure}[p]
  \centering
  \includegraphics[scale=0.8]{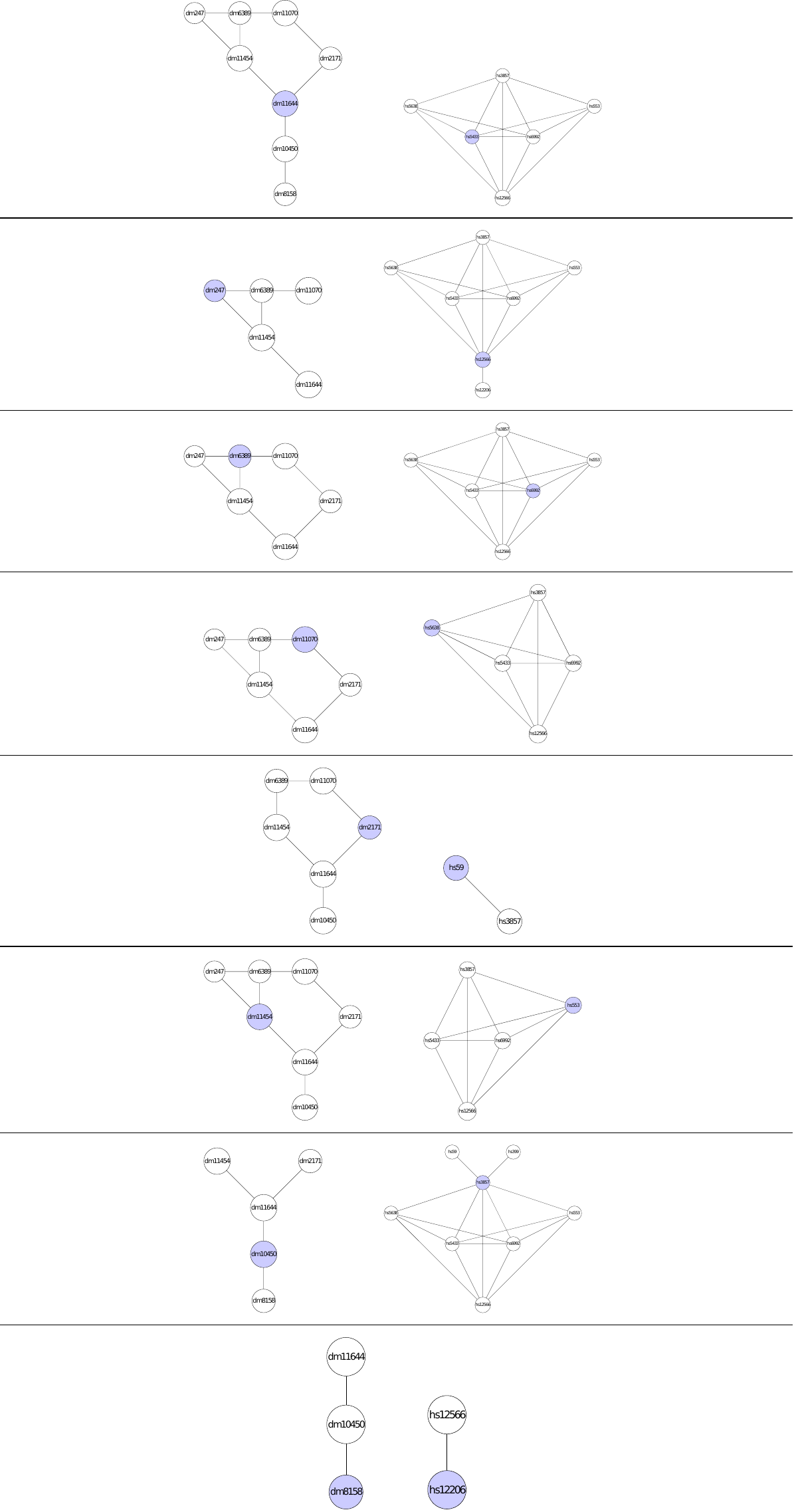}
\caption{\label{ClusterAssignment} This figure shows the final assignment between the clusters in 
Figure~2  produced by AligNet, which corresponds also to  Step 3. Each of the eight clusters obtained from $G$ is aligned to one, and only one, of the nine clusters obtained from $G'$. Hence, one cluster from $G'$ remains unmatched which is the second cluster  in the third row on the right in Figure~2.  In this figure,  we show  the clusters from $G$ on the left and its corresponding cluster image from $G'$ on the right.}
\end{figure}

\subsubsection*{Step 4. Local alignment of PPIN}

In this step,   AligNet  produces a local 
alignment between $G$ and $G'$  from the matching between $C(G)$ and 
$C(G')$ obtained in the previous step. The main idea is to define this alignment by merging the partial injective mappings $\eta_{u,u'}\in \mathcal{C}$. The problem is that these mappings may be inconsistent, because $C(G)$ and  $C(G')$ are 
overlapping clusterings. Indeed, it may happen that a node $w$ belongs to more than one cluster $C_u$,
and that   the corresponding mappings $\eta_{u,u'}\in \mathcal{C}$  send $w$ to different nodes in $G'$; conversely, 
for $w'$ belonging to multiple $C_{u'}$  mappings to  different nodes in $G$.

To overcome this problem, we consider a weighted bipartite hypergraph whose nodes are the nodes in $G$ and in $G'$, every mapping  $\eta_{u,u'}$ is a hyperarc with source its domain and target its image, and the weight of every hyperarc is the score $Score(\eta_{u,u'})$. Then, the solution of the weighted bipartite hypergraph assignment problem provides a well-defined local 
alignment of the input networks. However, in order to decrease the computation time of AligNet,  we do not consider all the mappings $\eta_{u,u'}$ together, but just a subset $\mathcal{R}$ of them that is recursively increased until all mapping  $\eta_{u,u'}$ have been considered. Thus,   AligNet  builds recursively a subset  $\mathcal{R}\subseteq \mathcal{C}$
of \emph{best-scored} alignments, by choosing, at each step,
a mapping  $\eta_{w_0,w_0'}\in \mathcal{C}$ with $w_0$ not belonging to the union of the domains of the mappings
$\eta_{w,w'}$ already in $\mathcal{R}$ and with maximum $Score(\eta_{w_0,w_0'})$ among all such mappings.
 AligNet iterates this procedure until  every node in $\bigcup_{\eta_{u,u'}\in \mathcal{C}} Dom\, \eta_{u,u'}$ belongs to the domain of some mapping in $\mathcal{R}$.
  In  Figure~5 
we give the subset $\mathcal{R}$ of $\mathcal{C}$ for the networks in our running example.

\begin{figure}[h]
\includegraphics[scale=0.8]{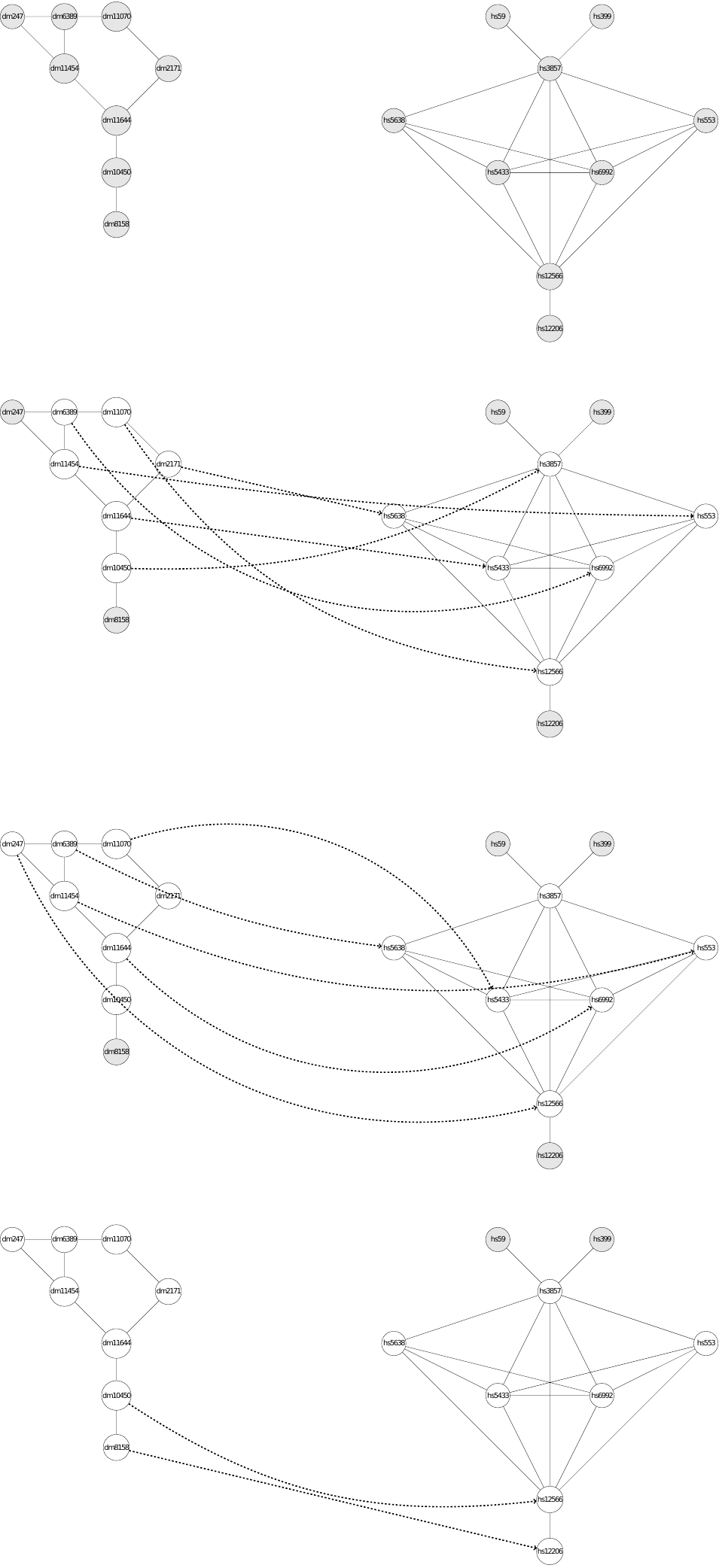}
\caption{\label{Recobriment} This figure shows  how AligNet constructs an appropriate set of 
alignments considered  to obtain a final local alignment. This corresponds to the Step 4 of our aligner.
First of all, a maximum score alignment between a pair of clusters  is chosen: in this case, this corresponds to the matching between the clusters in  Figure~3. Both clusters are shown in the second row of 
this figure. The shadowed nodes  are the nodes that are not aligned.
Next, a maximum score  alignment of a pair of clusters with source a cluster centered at a shadowed node is chosen: it turns out to be the one   in the second row   in  Figure~4 and it is shown in the third row in this figure. Finally, the last 
alignment to be included in the appropriate set of alignments must be the one with source cluster centered at the remaining shadowed node: this corresponds to the  alignment in the last row in Figure~4  shown in the bottom of this figure.  Notice that in the end, that is when we consider the three alignments together, there are four nodes in the source network with inconsistent assignments..}
\end{figure}

Now, consider the directed hypergraph $H$ with nodes $V\cup V'$
and hyperarcs the mappings $\eta_{u,u'}\in \mathcal{R}$: each $\eta_{u,u'}$ is understood as a hyperarc  with source its domain and target its image. Then, AligNet
obtains from this  hypergraph a local well-defined alignment  between $G$ and 
$G'$ as a solution of the corresponding weighted bipartite hypergraph assignment problem~\cite{borndoerfer.heismann:2015}.
Figure~6 shows the local alignment obtained from the hypergraph corresponding to Figure~5.

 \begin{figure}[ht!]
 \centering
 \hspace*{-1in}
\includegraphics[scale=0.8]{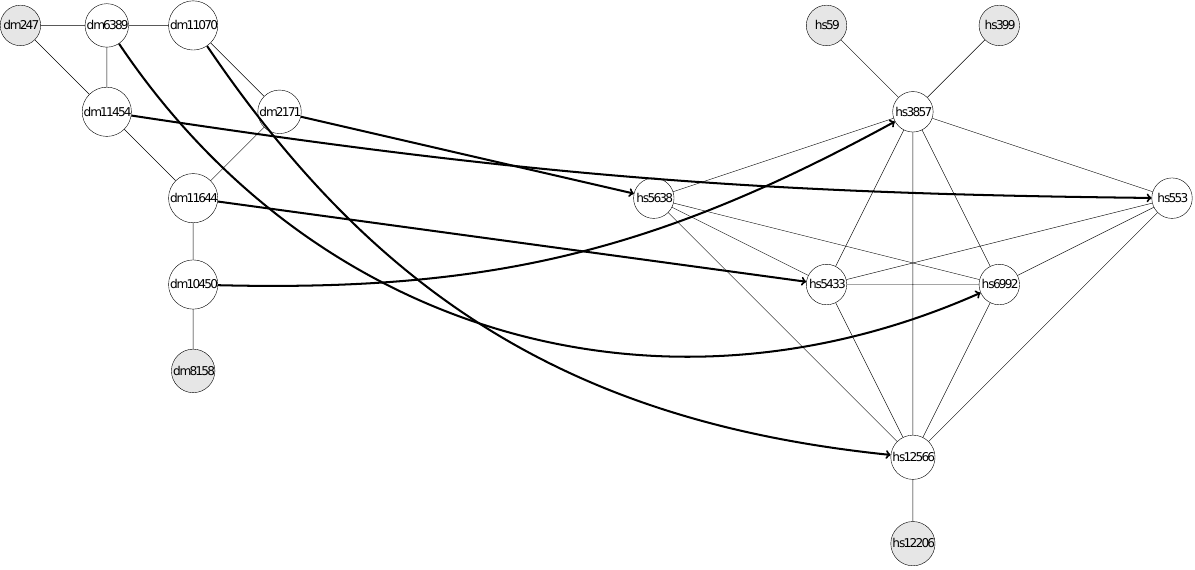}
\caption{\label{Global} This figure shows the local alignment of the original networks obtained by AligNet in its Step 4, once the inconsistent assignments have been solved. The coherent assignment of nodes is obtained as the solution to the weighted bipartite hypergraph assignment problem, for the hypergraph associated to the appropriate set of alignments described in Figure~5. In this case, the hypergraph has three hyperarcs, corresponding to the three alignments considered in the appropriate set of alignments..}
\end{figure}

\subsubsection*{Step 5. Global meaningful alignment of PPIN}

In order to extend the local alignment produced in the previous step, AligNet 
iterates the following  procedure: 

\begin{itemize}
\item It removes the nodes in $G$ and $G'$ that have already been aligned, and it recomputes the score of each alignment  $\eta_{u,u'}$  following the same definition as in Step 3, but only taking into account the remaining nodes in its domain and image. 

\item It computes a new optimal matching $\mathcal{C}$ between $C(G)$ and $C(G')$, as in step 3, but using as edges
those $\eta_{u,u'}$ whose updated score is positive, and weights these updated scores.

\item It computes a new set $\mathcal{R}$ of best-scored alignments $\eta_{u,u'}$ with $Score(\eta_{u,u'})>0$, as in step 4.

\item It defines a new directed hypergraph $H$ whose  nodes are the nodes in $V\cup V'$ not yet  aligned and hyperarcs  the mappings $\eta_{u,u'}$ in the new set $\mathcal{R}$, understood as hyperarcs with source the still unaligned nodes in their domain and target  the still unaligned nodes in their image. 

\item It computes a local alignment between unaligned nodes in $V$ and $V'$ by solving the weighted bipartite hypergraph assignment problem
for this hypergraph, and it adds this local alignment to the alignment obtained so far.
\end{itemize}

This procedure is iterated while there exist nodes not aligned belonging to the domain or the image of some alignment $\eta_{u,u'}$ with (updated) positive score: In Figure~7 we show the final global meaningful alignment obtained with  AligNet for the networks in our running example.

%
%
%Finally, if all the updated scores of the mappings $\eta_{u,u'}$ become 0 but  there are still unaligned
%proteins in both $G$ and $G'$, AligNet considers a complete bipartite graph where the nodes are the unaligned
%proteins and the weight of the edge between two nodes $u\in V$ and $u'\in 
%V'$  is the number of times that the node $u$ has been sent to the node $u'$ 
%by means of alignments $\eta_{w,w'}\in \mathcal{A}$. The matching between the remaining, unaligned proteins is obtained then as the result of the maximum  weighted bipartite matching algorithm applied to this  bipartite graph, and pasted to the alignment obtained so far. 

 \begin{figure}[htb!]
 \centering
 \hspace*{-1in}
\includegraphics[scale=0.8]{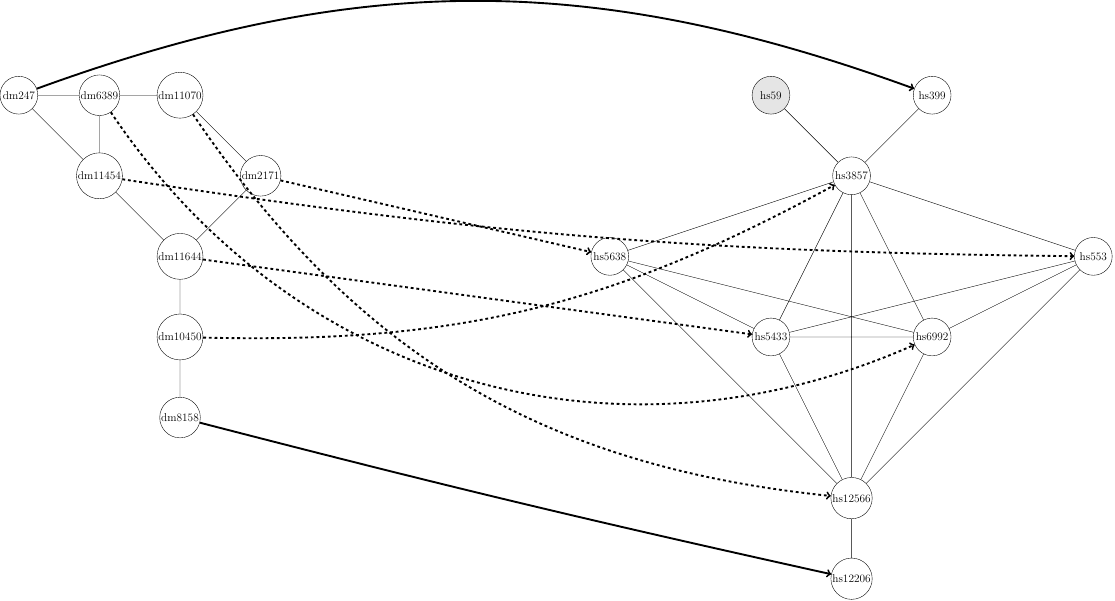}
\caption{\label{GlobalFinal} This figure shows the final global alignment of the original networks obtained by  AligNet.  Notice that, in  Step 5 of AligNet, the previous alignment is extended to a global one. In this case, there were two unmatched nodes in the source network in Figure~6 which are now assigned. The assignment of these two nodes is shown with solid arrows while with dashed arrows we show the already assigned nodes.}
\end{figure}

\subsection*{Evaluation of alignment quality}

Several methods have been proposed to evaluate the quality of an alignment and to compare the performance of PPIN aligners~\cite{Comparison,Ulign}. One of the handicaps when considering  biological data is that the true alignment is unknown and, therefore, to evaluate the alignment quality one cannot count the number of true and false positives and true and false negatives. However, several measures of alignment quality have been already proposed which are divided in two categories,  \emph{topological coherence} and \emph{biological coherence}.  The topological coherence measures evaluate the topological similarity of the aligned regions considering the \emph{edge correctness}, which is the percentage of conserved edges, the \emph{induced conserved substructure},    which is the percentage of preserved edges,  and the 
\emph{symmetric substructure score}, which is the percentage of preserved and conserved edges. The biological coherence measures evaluate the protein function similarities of the aligned proteins by considering  the  \emph{KEGG pathway annotation}, which measures the percentage of  proteins aligned to  proteins that participate in the same pathway, and the \emph{Gene Ontology annotation}, which measures the similarity between the GO terms of a protein and its image. In the comparison of eight recent aligners reported in~\cite{Ulign}, it is stated that there is a strong correlation between the three topological  coherence measures and also a strong correlation between the two biological coherence measures. Therefore, when evaluating the alignment quality, it is enough to consider one of the tree measures for topological coherence and one of the two biological coherence measures. However, there is a low correlation between the topological and the biological coherence measures. Thus, to evaluate the alignment quality, we considered both the edge correctness and the Gene Ontology annotation measures, which are defined as follows: 

Let $G=(V,E)$ and $G'=(V',E')$ be two PPIN such that 
$|V| \leq |V'|$. The \emph{edge 
correctness ratio} of a mapping $\mu : G \rightarrow G'$ is the ratio of 
the edges that are preserved  by $\mu$, and it is defined by
$$
EC(\mu)= \frac{\left|\big\{\{u,v\} \in E\mid \{\mu(u),\mu(v)\} \in E'\big\} \right|}{ 
min\{|E|, |E'|\}}.
$$

The \emph{functional coherence value}, or \emph{GO consistency}, of a mapping $\mu 
: G \rightarrow G'$  is defined as 
$$
FC(\mu)= \frac{\sum_{u\in V} FS(u,\mu(u))}{|V|},
$$
where the similarity score $FS$ is defined by 
 $$
 FS(u,u' ) = \frac{|GO(u) \cap GO(u')|}{|GO(u) 
\cup GO(u')|},
$$
 with $GO(u)$ and $GO(u')$  the sets of GO annotations of the 
proteins $u$ and $u'$, respectively.

\section*{Results and Discussion}

In this section we report the tests performed to assess the performance of AligNet. These tests have been  designed  taking into account the study and results reported  in \cite{Comparison} and \cite{Ulign}, where a comparison of algorithms for the pairwise  alignment of biological networks, and  in particular  of PPIN, is given. Considering the conclusions reported in both studies, the aligners proposed among all the consider aligners because they produce better alignments are NATALIE~\cite{NATALIE}, PINALOG~\cite{PINALOG},  SPINAL~\cite{SPINAL}, HubAlign~\cite{Hashemifar2014} and L-GRAAL~\cite{LGRAAL}.

 \subsection*{Network data}

We have considered  the same dataset used in~\cite{Comparison},  so that it makes sense to  compare the  results obtained by AligNet with the results reported therein. Thus, we have downloaded from the IsoBase database~\cite{IsoBase} the PPIN (version 1.0.2) of five  organisms: \emph{M.~musculus} (mus), \emph{C.~elegans} (cel), \emph{D.~melanogaster} (dme), 
\emph{S.~cerevisiae} (sce), and \emph{H.~sapiens} (hsa), and aligned each pair of them with AligNet.
The number of nodes and edges of these  PPIN 
are shown in  Table~\ref{nets/data}.  

\begin{table}%[ht]
\centering
\begin{tabular}{rrrrr}
  \hline
 & Nodes  & Edges (with loops) &  E (without loops) \\
  \hline
\emph{M. musculus} & 623 & 776 &  559 \\
\emph{C. elegans}   & 2995 & 8639 &  4827 \\
\emph{S. cerevisiae}  & 5524& 164718 &  82656 \\
\emph{D. melanogaster} & 7396 & 49467 &  24937 \\
\emph{H. sapiens} & 10403 & 105232 & 54654 \\
   \hline
\end{tabular}
\vspace{0.2cm}
\caption{\label{nets/data} This table shows the number of nodes and edges, with and without loops, of the five networks considered as input data in our tests.}
\end{table} 

\subsection*{Comparison to other aligners}

\subsubsection*{Quantitative analysis}

%Following the assessment of the existing global alignment algorithms carried on 
%in~\cite{Comparison, Ulign}, which suggests NATALIE~\cite{NATALIE}, PINALOG~\cite{PINALOG}, SPINAL~\cite{SPINAL}, HubAlign~\cite{}   and L-GRAAL~\cite{} as good aligners, we have chosen them to compare  AligNet  to.

In order to evaluate the quality of the alignments produced by  AligNet, and to compare it with that of  the aforementioned aligners, we have used both the edge correctness (EC) and 
the functional coherence (FC) metrics defined in the previous section.
% 
%Let $G=(V,E)$ and $G'=(V',E')$ be two PPINs such that 
%$|V| \leq |V'|$. The \emph{edge 
%correctness ratio} of a mapping $\mu : G \rightarrow G'$ is the ratio of 
%the edges that are preserved  by $\mu$, and it is defined by
%$$
%EC(\mu)= \frac{\left|\big\{\{u,v\} \in E\mid \{\mu(u),\mu(v)\} \in E'\big\} \right|}{ 
%|E|}.
%$$
%
%{\reversemarginpar\marginpar{\color{blue} Cesc: Què diu això? Segur que no voleu dir ``[\ldots] an edge correctness of 100\% should not be taken as a conclusive evidence of a correct network alignment, because \ldots??}}
%

As reported in \cite{Comparison}, an edge correctness ratio of 100\% should not be taken as a 
conclusive evidence of a correct network alignment, because it is always possible that two 
biologically unrelated edges had been mapped to each other. In addition, it is reported in \cite{Ulign} that the alignment of a dense network with a sparse network produces a low edge correctness ratio. We can observe that this situation is reflected for instance in the alignment  between \emph{S.~cerevisiae} and \emph{D.~melanogaster},  since there are more edges in the source network (sce) than in the target network (dme). 
%Actually, in this case L-GRAAL interchanges the order of the input networks, considering as source network the one with less edges. 
Thus, an edge correctness ratio of 20\% should be considered an evidence of an incorrect network alignment, only when the number of edges of the source network is smaller  than  the number of edges in the target network. Therefore, given the limitations of this  topological measure of alignment quality, measures of 
agreement derived from biological information are also popular in the 
literature. In particular, most papers on PPIN alignment 
make use of gene orthology annotations from the Gene Ontology (GO) database~\cite{GO} to measure alignment accuracy, by comparing the 
similarity of GO annotations between aligned proteins. Hence,  we have also used   the functional coherence value, or GO 
consistency, introduced in the previous section, to assess the quality of our alignments.

%The \emph{functional coherence value}, or \emph{GO consistency}, of a mapping $\mu 
%: G \rightarrow G'$  is defined as 
%$$
%FC(\mu)= \frac{\sum_{u\in V} FS(u,\mu(u))}{|V|},
%$$
%where the similarity score $FS$ is defined by 
% $$
% FS(u,u' ) = \frac{|GO(u) \cap GO(u')|}{|GO(u) 
%\cup GO(u')|},
%$$
% with $GO(u)$ and $GO(u')$  the sets of GO annotations of the 
%proteins $u$ and $u'$, respectively. 

In order to compare the EC scores and the FC scores of all the considered aligners for every pair of PPIN, we run all the aligners on the same pairs of input networks. However, as it has been already stated in previous studies of existing aligners~\cite{Comparison,Ulign},  some difficulties appear when trying to do this work. More precisely, there were computations that never stopped. This was the case of NATALIE.   L-GRAAL  matches the network with the smallest number of edges to the network with the largest number of edges, which means that  it interchanges the order of the input networks  in the case of the alignments between \emph{S.~cerevisiae} and \emph{D.~melanogaster} and between \emph{S.~cerevisiae} and \emph{H.~sapiens}. And also, for most of the aligners,  some parameters must  be fixed. We considered the parameters suggested by default in all the aligners whenever it was possible. With L-GRAAL, also a time limit or a maximum number of steps must be considered. We again decided to consider the parameters suggested by default in  its implementation. 

%as the limited time, the time consumed by the fastest computation when considering  the other aligners under the same pair of input networks. \red{Gabriel, com ho posaries això?}

In Table~\ref{ECscores} and   Table~\ref{FCscores} we report the results obtained with this test. We can observe there that the alignments of small networks with a low number of edges, such as \emph{M.~musculus},  produced alignments with  high EC scores, especially when the target network has a large number of edges. However, even in this case, the EC scores obtained with the aligners PINALOG and SPINAL are not  high. It is very surprising that SPINAL  preserved less than $10\%$ of the edges in the source network  in all the alignments, except  in the alignment  between \emph{M.~musculus} and  \emph{H.~sapiens}, only $24\%$ of the edges present in \emph{M.~musculus} were preserved. On the other hand, PINALOG obtained much more reasonable scores than SPINAL but in the best scenario, that is, when the alignment is between \emph{M.~musculus} and the others networks, PINALOG preserved less than $40\%$ of the edges in two alignments (\emph{M.~musculus} with \emph{C.~elegans} and \emph{D.~melanogaster}) while the other aligners preserved more than $60\%$ of the edges present in \emph{M.~musculus}. The best scores in this case are obtained by HubAlign, which preserves more than $80\%$ of the edges, followed by L-GRAAL that preserves the $70\%$ of the edges, and then  AligNet that preserves $60\%$ of the edges. However, we can also observe here that, when the number of edges in the source network increases, the EC scores decrease dramatically even in the case of HubAlign. When considering the alignments between \emph{S.~cervisiae} and \emph{D.~melanogaster} or \emph{H.~sapiens}, we observed that all aligners, even L-GRAAL that interchanges the input networks,  obtained less than $10\%$ of the edges matched in the target network. 

\begin{table}%[ht]
\centering\footnotesize
\begin{tabular}{rlrrrrr}
  \hline
  Net1 & Net2 & EC AligNet & EC HubAlign & EC L-GRAAL & EC PINALOG & EC SPINAL\\
  \hline 
   mus & cel & 0.58 & 0.81 & 0.79   & 0.34 & 0.01\\
   mus & sce & 0.65 & 0.97 & 0.68  & 0.56 & 0.05 \\
   mus & dme & 0.65 & 0.88 & 0.70& 0.30 & 0.03\\
   mus & hsa & 0.76 & 0.95 & 0.77 & 0.62 & 0.24 \\
  cel & sce & 0.24 & 0.83 & 0.38& 0.30 & 0.06\\
  cel & dme & 0.31 & 0.68 & 0.53 & 0.18 & 0.01\\
  cel & hsa & 0.31 & 0.77 & 0.43 &  0.23 & 0.01\\
   sce & dme & 0.03 & 0.01 & 0.08 &  0.19 & 0.03 \\
   sce & hsa & 0.04 & 0.03 & 0.13 & 0.19 & 0.04 \\
dme & hsa & 0.13 & 0.37 & 0.31 & 0.13& 0.01\\
   \hline
\end{tabular}
\caption{\label{ECscores} Edge Correctness scores obtained by the considered aligners.}
\end{table}

Therefore, we can conclude that the analysis of the obtained results  reveals that the EC score is only a measure of topological relation between the input networks. When the source network is smaller and has much less edges than the target network, then a high EC score should be expected. However, when the source network is similar to the target network or when it has a higher number of edges, then a very low EC score should be expected.  Overall, it clearly implies that another definition of alignment correctness must be considered.
 
% As it can be seen there, the ratio of conserved edges is not really high for any of the aligners but, even then, the best EC scores are obtained by  AligNet, with a 70\% of conserved edges in the best case. We can also observe in this figure that  the EC scores  obtained by SPINAL are very low. However, if we look at the FC scores, we observe that they are very low for all of the aligners, although SPINAL reaches the best scores for all but one of the alignments. We infer that this is due to the fact that SPINAL has a parameter to balance the weight between topological and sequence similarities. We  fixed the parameter to $0.7$, as suggested in \cite{Comparison}, so that sequence similarity has more weight than topological similarity. Nevertheless, the FC score is around 20\% of  GO consistency in all of the alignments.\red{ ***} Therefore, from the results obtained with this test we can conclude that AligNet obtains a higher edge correctness ratio than the other aligners but, 

 As far as the functional coherency goes, we report the obtained results in Table~\ref{FCscores}. We can observe there that all the aligners obtained a very low score. However, we cannot conclude that all aligners have a low biological coherence, because it is not clear if the low value is due to the alignment itself or to the measure of biological coherence. Therefore, we tried to obtain the maximum value of the FC score that can be expected for every pair of networks. To obtain this value we performed the following test: for every pair of input networks, we considered a complete bipartite graph where  the nodes are the proteins in the two input networks, the edges are all  protein pairs consisting on a protein  in the source network and a protein in the target network,  and the weight of each edge is the FC score of the corresponding pair of proteins.  Then, we obtain the maximum FC score, 
  $FC_{max}$, as the FC score of the solution to the maximum weighted bipartite matching problem.   Hence, for every pair of networks, we can compare the FC score obtained for each aligner with the maximum  score. We define the \emph{relative biological coherence} as the ratio between the FC score and the  $FC_{max}$.  That is, $FC_{rel}=\frac{FC}{FC_{max}}$. Now, if we look at the results presented in  
  Table~\ref{FCscores}, we can observe that the values of $FC_{max}$ range from $0.19$ to $0.26$ and only in the alignment between \emph{M.~musculus} and \emph{H.~sapiens} we obtain that $FC_{max}= 0.54$, which means that, in average in most of the alignments, the best alignment from the functional coherence point of view, maps correctly only $20\%$ of  the proteins. Considering now  the results obtained by the different aligners, we can observe that the order from the highest to the lowest scores is almost the opposite to the order obtained when considering the EC scores. That is, the best scores are achieved by SPINAL, followed by PINALOG, AligNet, HubAlign and L-GRAAL. However, the results obtained even with SPINAL is that only $10\%$ of the proteins is matched correctly, considering the GO term measure of biological coherence.  In addition, the best possible alignment would be able to align correctly only $20\%$ of the proteins. We discard  the hypothesis that the GO term measure is low due to the lack of GO terms, since, as we show in Table~\ref{GOs}, for every pair of networks, $90\%$ of protein pairs have their GO terms annotated, and there is no  correlation between the number of annotated GO terms and FC scores. 
Therefore, since the FC scores are not conclusive of meaningful biological alignments, we performed an additional test explained below.  

\begin{table}[ht]
\centering\footnotesize
\begin{tabular}{rllrrrrlll}
  \hline
 Net1 & Net2 & $FC_{max}$ & AligNet     & HubAlign & L-GRAAL  & PINALOG & SPINAL \\ 
  & &  & FC -- $FC_{rel}$     &  FC -- $FC_{rel}$ & FC -- $FC_{rel}$ & FC -- $FC_{rel}$ &  FC -- $FC_{rel}$ & \\
  \hline
 mus & cel & 0.21 & 0.06 --  0.26 & 0.04 --  0.21 & 0.03 -- 0.17 & 0.10 --  0.50 & 0.12 --  0.60 \\
mus & sce & 0.24 & 0.08 --  0.33 & 0.07 --  0.30  & 0.04 -- 0.18 & 0.12 --     0.50 & 0.15 --  0.64 \\
 mus & dme & 0.19 & 0.05 -- 0.24 & 0.03 -- 0.16  & 0.03 -- 0.13 & 0.07   -- 0.42  & 0.06  --    0.33 \\
 mus & hsa & 0.54 & 0.23 -- 0.42 &  0.26 -- 0.49  & 0.10  -- 0.18 & 0.48  -- 0.90 &  0.10 --   0.20  \\
 cel & sce & 0.20 & 0.06 -- 0.33 & 0.03 -- 0.14     & 0.04  -- 0.21& 0.13  --   0.70 &  0.19 --  0.99  \\
 cel & dme & 0.23 & 0.04 -- 0.18 & 0.02 -- 0.07   & 0.02  -- 0.09 & 0.09  --   0.42 & 0.09 -- 0.42\\
 cel & hsa & 0.24 & 0.04 -- 0.17 & 0.02 --  0.08   & 0.03 -- 0.13 & 0.08  -- 0.35 &  0.08 --  0.36 \\
 sce & dme & 0.24 & 0.05 -- 0.19 & 0.07 -- 0.30 & 0.02  -- 0.07& 0.07 --   0.31 &   0.10 --  0.43 \\
 sce & hsa & 0.26 & 0.06 -- 0.24 & 0.08  -- 0.30 & 0.02  -- 0.07& 0.09 --  0.29 &  0.11 --  0.45 \\
 dme & hsa & 0.20 & 0.04 -- 0.18 & 0.02 -- 0.11 & 0.02 -- 0.08 & 0.09 --   0.41 &  0.08 --  0.43 \\
 \hline
\end{tabular}
\caption{\label{FCscores} Functional  Coherence scores obtained by the considered aligners.}
\end{table} 

\begin{table}[ht]
\centering\footnotesize
\begin{tabular}{rllrrrrr}
  \hline
 Net1 & Net2 & $FC_{max}$ & NotAvailableGOpairs & AvailableGOpairs & Percentage \\
  \hline
   mus & cel & 0.21 &   1 & 598 & 99.83 \\
   mus & sce & 0.24 &  0 & 599 & 100.00 \\
    mus & dme & 0.19 &  2 & 597 & 99.67 \\
  mus & hsa & 0.54 &   1 & 598 & 99.83 \\
    cel & sce & 0.20 &   20 & 2954 & 99.33 \\
   cel & dme & 0.23 &  256 & 2718 & 91.39 \\
   cel & hsa & 0.24 &  102 & 2872 & 96.57 \\
   sce & dme & 0.24 &   45 & 5478 & 99.19 \\
   sce & hsa & 0.26 &   14 & 5509 & 99.75 \\
    dme & hsa & 0.20 & 0.19 & 7126 & 96.47 \\
   \hline
\end{tabular}
\caption{\label{GOs} This table shows the relation between available GO pairs for every pair of networks.}
\end{table} 

 \subsubsection*{Qualitative analysis}
 
As stated in  \cite{Comparison}, the evaluation of any aligner should be done considering also their quality and not only the quantity, that is, the accuracy of the alignment and not only the number of preserved edges or GO terms. The accuracy of any aligner is easily tested when there is a gold standard to compare to. With this idea in mind, we have considered the following test to study the quality of our alignments in contrast with the quality of  the alignments obtained by PINALOG, HubAlign and L- GRAAL. Since the results obtained by SPINAL in the previous test were not convincing, we do not consider it in the next test.
   
\subsubsection*{Protein complex prediction}
In order to test  the behavior of  AligNet in the alignment of protein complexes,  we also performed the protein complex prediction test reported in~\cite{PINALOG} using PINALOG.  Following the procedure explained therein, we considered  the database MIPS CORUM \cite{CORUM} for the human protein complexes and, as a gold standard for the yeast complexes, we considered the  information available in \cite{SCEComplex}.  In addition, we considered the functional information available in MIPS CORUM for the human complexes and in
MIPS FunCat \cite{FunCat} for the yeast complexes.
Then,  we  considered the overlapping score of  complexes introduced  in~\cite{PINALOG}, and we defined a functional coherence value for the alignment of protein complexes, as explained below.

Let $G=(V,E)$ and $G'=(V',E')$ be two PPIN such that 
$|V| \leq |V'|$, let $\mu: G \rightarrow G' $ be a mapping, and let  $c\subseteq V$ and $c'\subseteq V'$ 
be two protein complexes in $G$ and $G'$, respectively. The \emph{overlapping 
score of $c$ and $c'$} is defined as
$$
OS(c, c') = \frac{ | \{ u\in c \; | \; \mu(u) \in c' \}|}{\min ( |c|, |c'| ) 
}.
$$

To define a 
functional coherence value for the protein complex alignment, every protein complex $c$ in $G$
is first mapped to a protein complex 
$c'$ in $G'$ provided that $OS(c,c')$ lies over a threshold, fixed at $0.2$ as done in~\cite{PINALOG}. Next, for every complex $c$ in $G$ we consider the complex  $c'$ in $G'$ such that $OS(c,c')$ is maximum. 

As it was the case with the edge correctness ratio, an overlapping score of pairs of complexes  of 100\% need not be evidence of a correct network alignment, 
because every protein complex is supposed to develop several biological functions, 
and the alignment may establish a correspondence between two complexes that are completely unrelated  from the point of view of their function.

The main point here is that the aim of the alignment should 
be clearly stated. If it only aims at 
matching similar  topological substructures of the networks, in order to detect
those substructures that appear in both networks, then maximizing the sum of the overlapping 
score of pairs of complexes may be a suitable goal.
 However, if the alignment searches for pairs of proteins that share 
 biological functions, then only those complexes with a common function 
should  be matched. Since the main application of PPIN alignment is to infer 
biological functions of proteins  and protein complexes,
 it is very important that the alignment does not match biologically unrelated
 complexes. Therefore, we define the complex functional coherence 
 of an alignment between PPIN as follows.  First, a pair of two complexes, one in each network, is  said to be 
\emph{coherent}  if they share some biological function; otherwise,  the pair is incoherent.
 Then, the \emph{complex functional coherence value}  (CFC) of the alignment  is defined by  the complex alignment precision,
 that is, the ratio of complexes that are aligned correctly with respect to the aligned complexes. If we denote by $CP$ the number of coherent pairs and by $NCP$ the number of incoherent pairs, then $CFC=\frac{CP}{CP + NCP}\times 100 $.

In Table~\ref{Complexes} we show the results obtained by all the aligners. We can observe there that AligNet does not align $1269$ complexes and produces $377$ incoherent pairs and $128$ coherent pairs. HubAlign does not align $1154$ complexes and produces $589$ incoherent pairs and  $31$ coherent pairs. PINALOG does not align $945$ complexes and produces $626$ incoherent pairs and $203$ coherent pairs. Finally, L-GRAAL does not align $996$ complexes and produces $741$ incoherent pairs and $31$ coherent pairs. Thus, the CFC value obtained in the alignment produced by AligNet is 25.34, which means that AligNet aligns correctly $25\%$ of the assigned complexes. Very close to AligNet is the CFC value in the alignment produced by PINALOG, which is $24.48$. However, the CFC values obtained in the alignments produced by L-GRAAL and HubAlign are lower. They are  $4.75$ and $5$ respectively, which means that they align correctly $5\%$ of the matched complexes.

\begin{table}%[ht]
\centering
\begin{tabular}{rrrrr}
  \hline
 & AligNet & HubAlign & PINALOG & L-GRAAL    \\   
   \hline
NotAssigned & 1269 & 1154 & 945 & 996 \\
 NotCoherent & 377.00 & 589.00 & 626.00 & 741 \\
  Coherent & 128 & 31 & 203 & 37  \\
  CFC & 25.34 & 5 & 24.48 & 4.75 \\
   \hline
\end{tabular}
\caption{\label{Complexes} This table shows the number of complexes that are not assigned/assigned correctly and assigned incorrectly.}
\end{table} 

In order to analyze the results  obtained by  AligNet in contrast to  the others aligners,  we also considered the following metrics: to contrast AligNet versus another aligner, for instance HubAlign,  we first count the number of complexes that are not aligned either by AligNet nor by HubAlign. Then, for those complexes that are aligned by AligNet and are not aligned by HubAlign, we count the number of coherent and incoherent pairs. Conversely, for those complexes that are not aligned by AligNet and are aligned by HubAlign, we count the number of coherent and incoherent pairs. We show these results in Table~\ref{AligNet/HubAlign}. We can see there that there are $891$ complexes that are not aligned neither by AligNet nor by HubAlign. This means that $77\%$ of the complexes that are not aligned by AligNet are also not aligned by HubAlign. With respect to the remaining $263$ complexes that are aligned by AligNet but not by HubAlign, $88$ are correctly aligned (coherent pairs) and $175$ are incorrectly aligned. On the other hand, there are $378$ complexes that are aligned by HubAlign but not by AligNet, from those, $21$ are correctly aligned and $357$ produced incoherent pairs. Therefore, we can conclude that HubAlign assigned more complexes than AligNet, but, when we look in detail to the complexes alignment, AligNet was able to align correctly $88$ complexes that HubAlign did not align. HubAlign aligned correctly only $21$ complexes that AligNet did not align. In addition, HubAlign aligned incorrectly $357$ complexes that AligNet did not align and AligNet aligned incorrectly $ 175$ complexes that HugAlign did not align. This means that AligNet achieves a higher precision in complex alignment than HubAlign.

\begin{table}%[ht]
\centering
\begin{tabular}{rrlrl}
  \hline
& AligNet vs HubAlign & Ratio &  HubAling vs AligNet & Ratio \\
  \hline
Not Assigned &  891 & 77.21 & 891 & 70.21     \\
  Not Coherent &  175 & 15.16 & 357 & 28.13   \\ 
  Coherent &  88  &7.63 & 21 &  1.65    \\
   \hline
\end{tabular}
\caption{\label{AligNet/HubAlign} This table shows  how AligNet assigned the complexes that are not assigned by HubAlign and conversely.}
\end{table} 

Concerning the results shown in Table~\ref{AligNet/L-GRAAL},  when we contrast AligNet with L-GRAAL, we obtain a similar result to the contrast of AligNet with HubAlign. Again, AligNet obtains a higher precision than L-GRAAL since it aligned correctly $66$ complexes and incorrectly $167$, while L-GRAAL aligned correctly $29$ complexes and incorrectly $477$. 

\begin{table}%[H]
\centering
\begin{tabular}{rrlrl}
  \hline
 & AligNet vs L-GRAAL & Ratio &  L-GRAAL vs AligNet &Ratio \\
  \hline
Not Assigned &  763 & 76.61 & 763 &60.13 \\
  Not Coherent & 167 & 16.77 & 477 &37.59 \\
  Coherent & 66 &6.63 & 29  &2.29 \\
   \hline
\end{tabular}
\caption{\label{AligNet/L-GRAAL} This table shows  how AlignNet assigned the complexes that are not assigned by HubAlign and conversely.}
\end{table}

However, the results obtained when we contrast AligNet with PINALOG show that, indeed, they obtain a similar precision in complex alignment (See Table~\ref{AligNet/PINALOG}). We obtain that AligNet aligned correctly $25$ complexes and incorrectly $105$ while PINALOG aligned correctly $79$ complexes and incorrectly $375$. Thus, the ratio of coherent pairs over assigned complexes is similar, $0.19$ for  AligNet and $0.17$ for PINALOG. Therefore, the only differences that can be observe between these aligners is that PINALOG aligns more complexes than AligNet. If it aligns more complexes, and its precision is slightly the same as the precision of AligNet, then PINALOG has more incorrect alignments than AligNet and also more correct alignments. In this sense, AligNet is a more conservative aligner than PINALOG, although its precision is slightly higher than the precision of PINALOG.

\begin{table}%[ht]
\centering
\begin{tabular}{rrlrl}
  \hline
 & AlignNet vs PINALOG & Ratio  & PINALOG vs AlignNet  & Ratio \\
  \hline
Not Assigned  & 815.00 & 86.24 & 815.00 & 64.22 \\
  Not Coherent &  105 & 11.11 & 375 & 29.55  \\
  Coherent & 25 & 2.65 & 79 & 6.23  \\
   \hline
\end{tabular}
\caption{\label{AligNet/PINALOG} This table shows  how AligNet assigned the complexes that are not assigned by PINALOG and conversely.}
\end{table} 

\begin{figure}[ht]
 \centering
 \hspace*{-1in}
\includegraphics[scale=0.6]{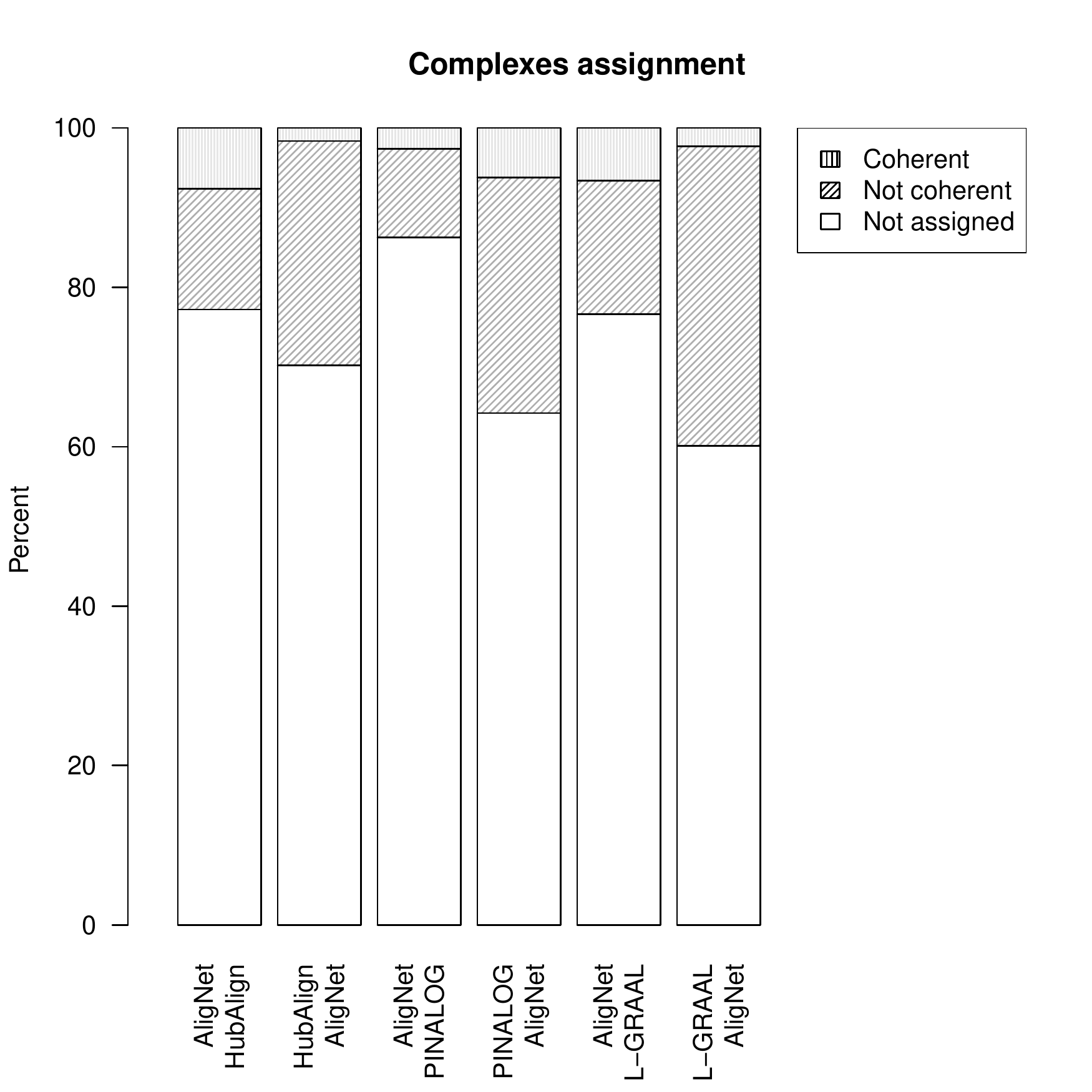}
\caption{\label{boxplotCoherence} This figure shows the results  in the protein complexes test obtained by AligNet in contrast to the others aligners, when we consider the alignment of complexes between S. cerevisiae and H. sapiens. We show the  proportion obtained by AligNet of  coherent, not coherent and not assigned complexes when the other aligners do not assign complexes, and conversely. Thus, the first bar shows the proportion between coherent, not coherent and not assigned complexes by AligNet when HubAlign does not assigned. Conversely, the second bar shows the proportion between coherent, not coherent and not assigned complexes by HubAlign when AligNet does not assigned.}
\end{figure}

We present in  Figure~\ref{boxplotCoherence} a visualization of the obtained results when we contrasted AligNet with the other aligners. We present there the ratio of unaligned complexes, correctly aligned complexes (coherent pairs) and incorrectly aligned complexes (incoherent pairs). We can observe that HubAlign versus AligNet (second bar from the left) as well as L-GRAAL versus AligNet (first bar from the right) obtain a higher proportion of incoherent pairs and a lower proportion of coherent pairs. In contrast, AligNet versus PINALOG and PINALOG verus AligNet (the two bars in the center) obtain a similar proportion of correctly and incorrectly aligned pairs. 

As a result of the comparison between the aligners, we obtain again, as it was the case in \cite{Comparison} and \cite{Ulign}, that the agreement of the alignments obtained by different aligners is vey low.  The majority of the global aligners achieve a high node coverage, meaning that the average of assigned nodes in the source network is high, but all of them obtain a very low biological coherence value. With respect to the topological coherence value, some aligners are able to obtain a high score but it is always associated with a biological coherence score. Overall, we can conclude that AligNet is the aligner that obtains a better balance between topological coherency (it preserves $60\%$ of the edges) and functional coherency (relative function coherency values between $20\%$ and $40\%$ and the highest complex functional coherency score, $25.34$), followed by PINALOG, which obtains similar functional coherency scores than AligNet and a bit lower topological coherency scores. However, HubAlign and L-GRAAL obtain high topological coherency scores but low functional coherency values. On the other hand, SPINAL surprisingly obtains a very low topological coherency value. Thus, if the purpose of the alignment is to align correctly, in the biological function sense, we propose to adopt AligNet since it is the most precise. However, if the purpose of the alignment is to find some  topological similarity, then we propose to adopt HubAlign.

\subsection*{Aligners analysis}

%Usually, the efficiency or running time of an algorithm is stated as a function relating the input length to the number of steps (time complexity) or storage locations (space complexity). Thus, 

In order to study the  efficiency of the considered aligners, we take into account their running time and memory space needed to perform an alignment. We run our implementation of AligNet on a server with 4 processors at 2.6 GHz and 20 GB of RAM and we also run the latest implementation of NATALIE (downloaded from \url{http://www.mi.fu-berlin.de/w/LiSA/Natalie}), PINALOG (downloaded from \url{http://www.sbg.bio.ic.ac.uk/~PINALOG/}),  SPINAL (downloaded from \url{http://code.google.com/p/spinal/}), HubAlign (downloaded from \url{http://ttic.uchicago.edu/∼hashemifar/software/HubAlign.zip}) and L-GRAAL (downloaded from \url{ http://bio-nets.doc.ic.ac.uk/L-GRAAL/}). 

\begin{figure}[ht]
 \centering
\hspace*{-1in}
\includegraphics[scale=0.6]{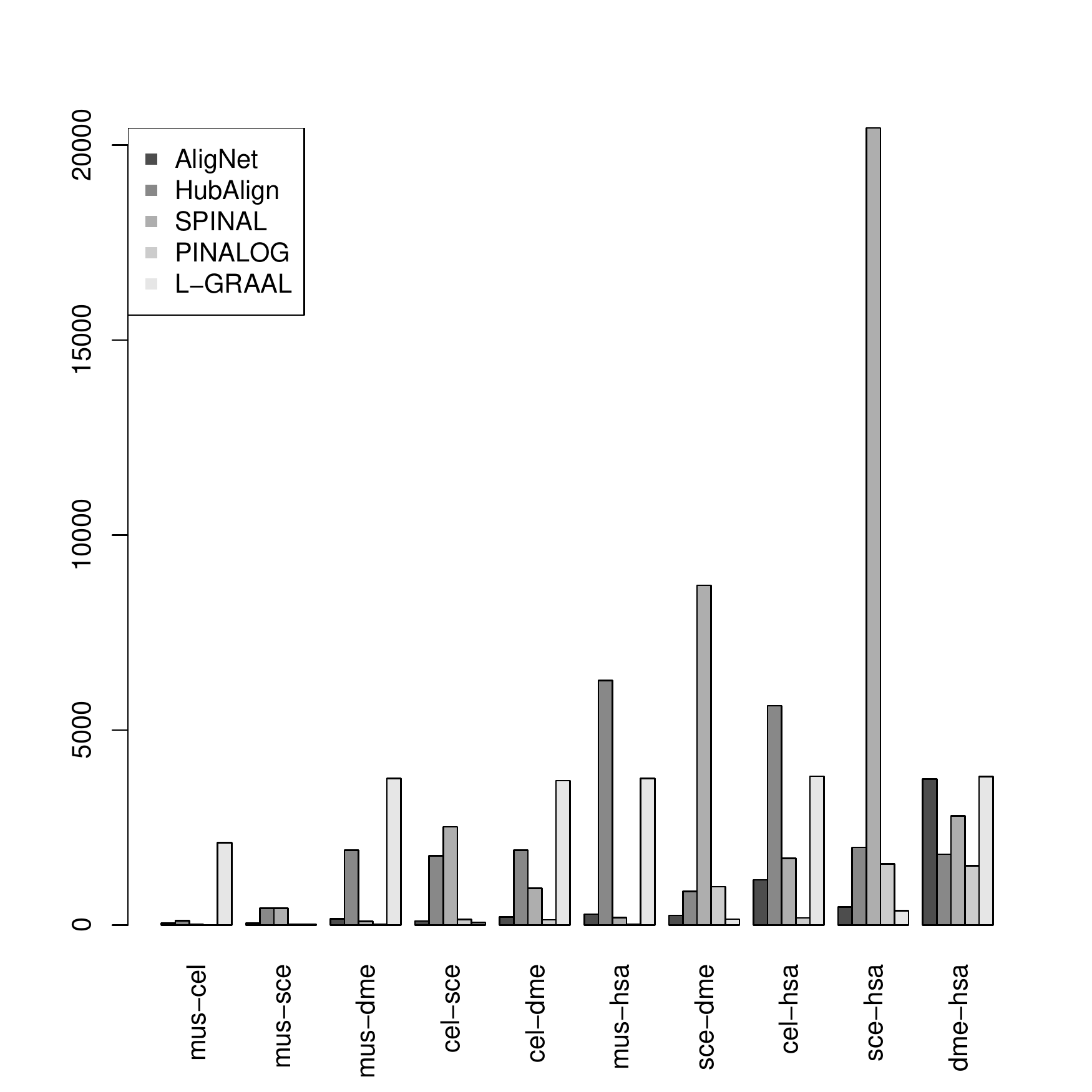}
\caption{\label{RunningTimes1} This figure shows the running times (in seconds) we obtained when we performed all the alignments for every pair of the considered networks. In the figure we present the results obtained with the aligners AligNet,  PINALOG, SPINAL, HubAlign and L-GRAAL.}
\end{figure}

\begin{figure}[ht]
 \centering
 \hspace*{-1in}
\includegraphics[scale=0.6]{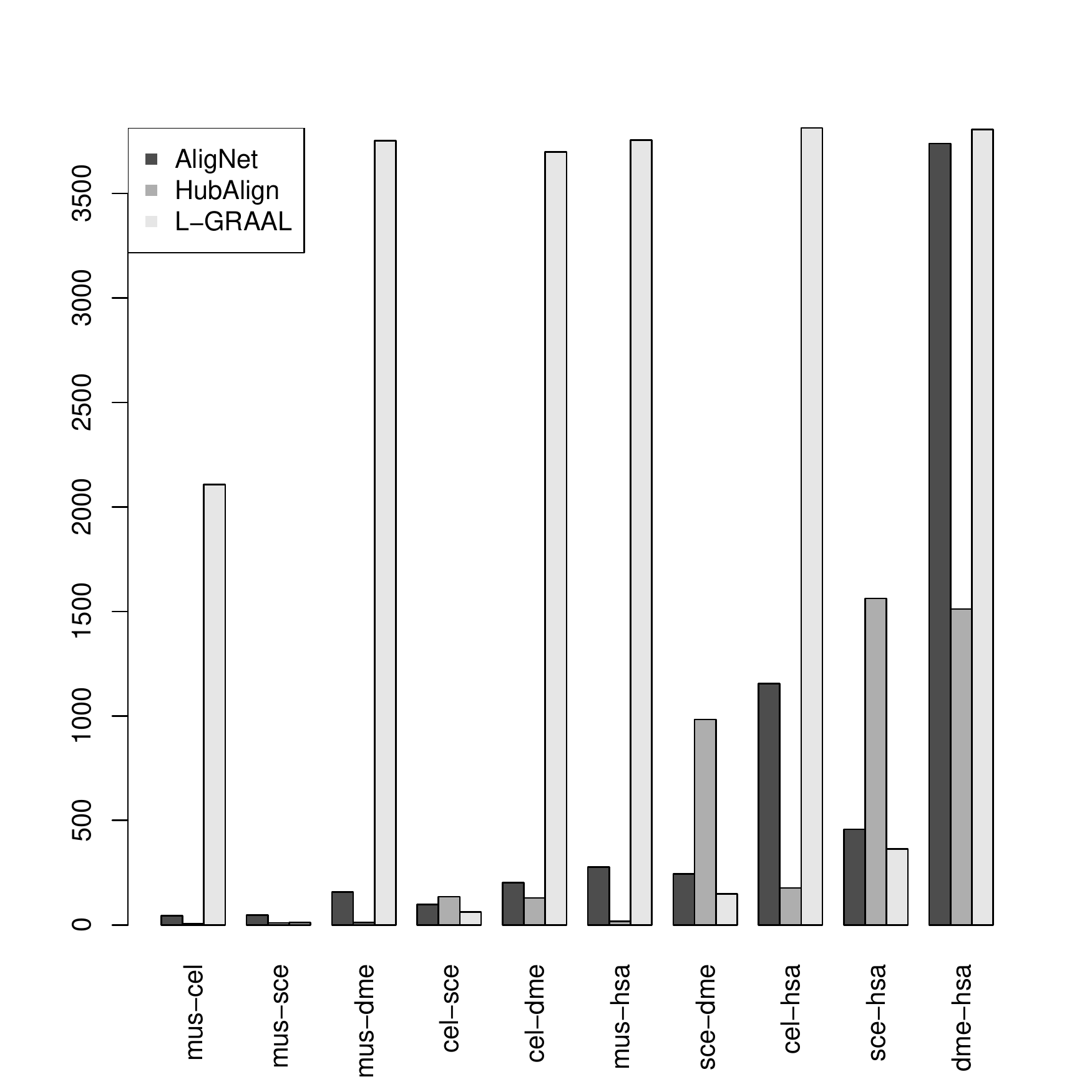}
\caption{\label{RunningTimes2} This figure shows the  same information presented in Figure~9 considering only the aligners AligNet, HubAlign and L-GRAAL.}
\end{figure}

As we already explained in the Background section, one of the weak points of PPIN aligners is either their running time or the memory space they use. Indeed, although NATALIE was suggested as a good aligner, it could not even align the two smallest networks, \emph{C.~elegans} and \emph{D.~melanogaster}, on a computer with 64 GB of RAM. With respect to PINALOG, SPINAL, HubAlign and L-GRAAL, we were able to complete all the alignments and we show their running times in  Table~\ref{nets/time}. In order to visualize their running times,  we also show the running times of every finished computation for each aligner in Figure~\ref{RunningTimes1}. We can observe there that SPINAL is, with a big difference, the slowest one to compute the alignments between \emph{H.~sapiens} and \emph{S.~cerevisiae},  and also between \emph{D.~melanogaster} and \emph{S.~cerevisiae}. In addition, PINALOG is the slowest one, also with a big difference, to compute the alignment between \emph{C.~elegans} and \emph{H.~sapiens}, as well as the alignment between \emph{H.~sapiens} and \emph{M.~musculus}. We can also observe that  AligNet is considerably faster than PINALOG and SPINAL, with a running time of less than a thousand seconds in most of the alignments. Only in one computation, the alignment between  \emph{D.~melanogaster} and  \emph{H.~sapiens}, AligNet is  slower than PINALOG and SPINAL  with a difference of less than two thousand seconds. However,  it is difficult to see the running times in some alignments because SPINAL needed more than $20,000$ seconds for the alignment between \emph{S.~cerevisiae} and \emph{H.~sapiens}. Thus, in order to visualize the results in the cases where the aligners consumed less than $3,500$ seconds, we decided to remove SPINAL and PINALOG  and in Figure~\ref{RunningTimes2} we show agin the results considering only AligNet, HubAlign and L-GRAAL.  We can observe there that L-GRAAL is the aligner that consumed more time in most of the computations. Concerning HubAlign and  AligNet, HugAlign is faster except in the alignments between \emph{C.~elegans} and \emph{S.~cerevisiae}  and also \emph{S. Cerevisiae} and \emph{D. melanogaster}.

\begin{table}
\begin{center}
\begin{tabular}{lll} \hline
\textbf{Alignment} &  & \textbf{Time} \\  \hline
cel-dme & Compute Matrices & 18.3\\
& Overlapping Clustering & 47.5\\
& Clusters Alignment and Assignment& 105.45\\
& Global Alignment & 113.173\\
& \textbf{Total} & \textbf{331.098}\\ \hline
cel-hsa & Compute Matrices & 49.68 \\
& Overlapping Clustering & 79.28 \\
& Clusters Alignment & 109.198 \\
& Global Alignment & 215.246 \\
& \textbf{Total} & \textbf{555.328} \\ \hline
cel-mmu & Compute Matrices & 3.914 \\
& Overlapping Clustering & 5.436 \\
& Clusters Alignment & 17.556 \\
& Global Alignment & 1.774 \\
& \textbf{Total} & \textbf{56.381} \\ \hline
cel-sce & Compute Matrices & 14.422 \\
& Overlapping Clustering & 28.59 \\
& Clusters Alignment & 29.663 \\
& Global Alignment & 42.676 \\
& \textbf{Total} & \textbf{147.788} \\ \hline
dme-hsa & Compute matrices & 293.08 \\
& Overlapping Clustering & 125.215 \\
& Clusters Alignment & 277.877 \\
& Global Alignment & 1195.752 \\
& \textbf{Total} & \textbf{2108.493} \\ \hline
dme-mmu & Compute Matrices & 10.684 \\
& Overlapping Clustering & 24.722 \\
& Clusters Alignment & 72.988 \\
& Global Alignment & 10.638 \\
& \textbf{Total} & \textbf{176.482} \\ \hline
dme-sce & Compute Matrices & 171.263 \\
& Overlapping Clustering & 60.953 \\
& Clusters Alignment & 66.736 \\
& Global Alignment & 192.9 \\
& \textbf{Total} & \textbf{542.181} \\ \hline
hsa-mmu & Compute Matrices & 21.392 \\
& Overlapping Clustering & 46.998 \\
& Clusters Alignment & 94.407 \\
& Global Alignment & 14.947 \\
& \textbf{Total} & \textbf{288.964} \\ \hline
hsa-sce & Compute Matrices & 56.28 \\
& Overlapping Clustering & 101 \\
& Clusters Alignment & 380.864 \\
& Global Alignment & 424.039 \\
& \textbf{Total} & \textbf{1066.908} \\  \hline
mmu-sce & Compute Matrices & 7.792 \\
& Overlapping Clustering & 14 \\
& Clusters Alignment & 13.107 \\
& Global Alignment & 1.567 \\
& \textbf{Total} & \textbf{61.630} \\ \hline
\end{tabular}
\caption{\label{nets/time} AligNet running times in seconds.}
\end{center}
\end{table}

\begin{figure}[ht]
 \centering
 \hspace*{-1in}
\includegraphics[scale=0.6]{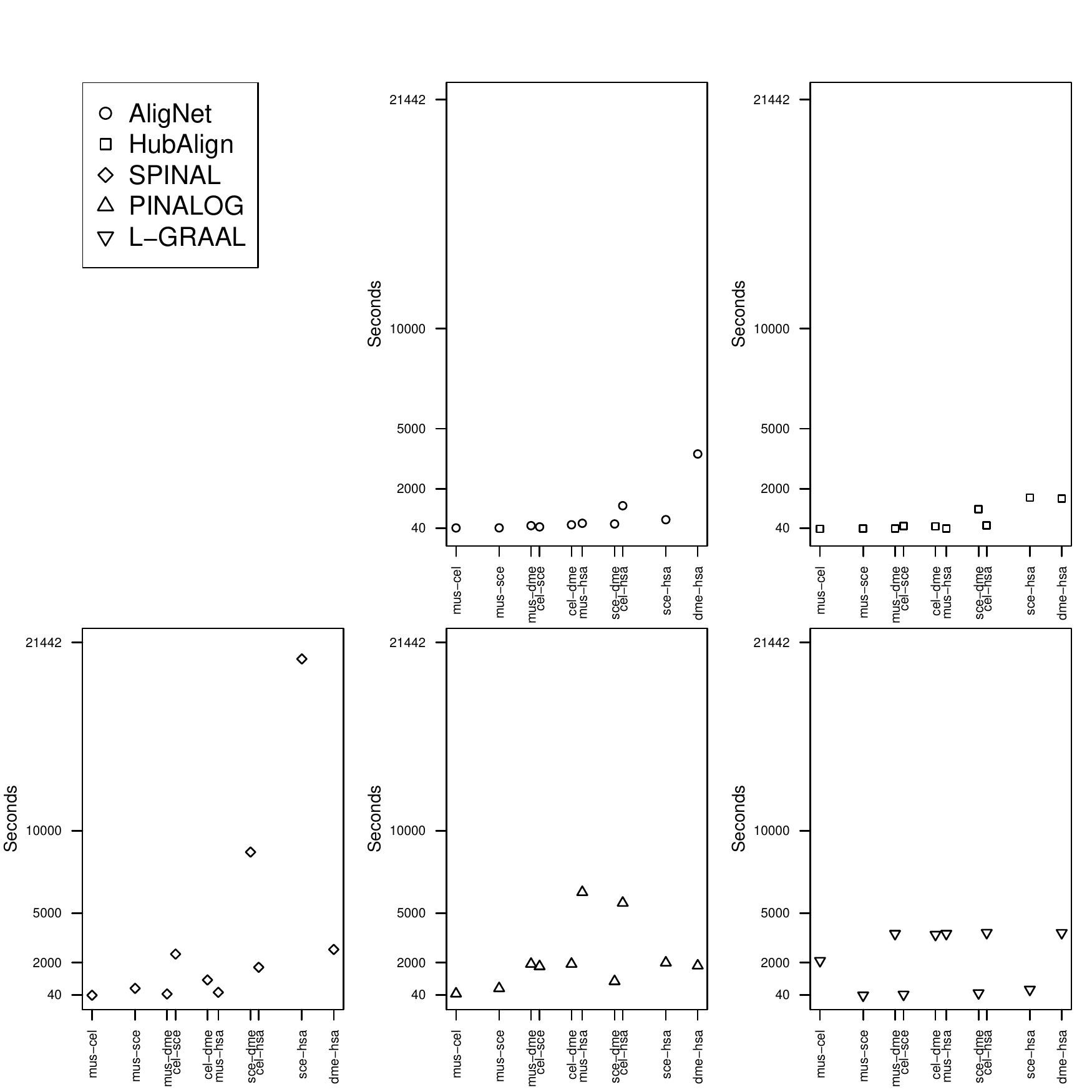}
\caption{\label{RunningTimes3} This figure shows the relation between the time needed to obtain every alignment and the size of the input networks with the aligners  AligNet,  PINALOG, SPINAL, HubAlign and L-GRAAL.}
\end{figure}

Furthermore, we show in Figure~\ref{RunningTimes3} the relation between network size and running time for all of the computations with each of the aligners. The size of a network pair is the sum of their nodes. Thus,  the network pairs in the diagrams are  positioned  in increasing order.  
A perfect aligner, from the efficiency point of view, should present a  linear relation between  the size of the network pair and the consumed time. From top left to bottom right, we show the results for the aligners AligNet, HubAlign, SPINAL, PINALOG, and L-GRAAL. 
We can observe  that HubAlign and AligNet present a clear relation between computation time and size of the input networks. However, this is not the case of PINALOG, SPINAL and L-GRAAL. It should be noticed here, that L-GRAAL has a step parameter which may force to stop the computation.

\section*{Conclusions}

In this paper we present AligNet, a new method and software tool for the pairwise global alignment of PPIN aimed to produce  
biologically  meaningful alignments  by achieving a good balance between structural matching and protein function conservation. AligNet is a parameter-free algorithm that, given two PPIN, produces a consistent alignment from the smaller network, in terms of number of nodes, to the larger network. In order to assess the correctness of our AligNet aligner, we have evaluated the quality of the alignments  obtained with AligNet and with the best aligners established in~\cite{Comparison,Ulign}, namely: PINALOG, SPINAL, HubAlign, and L-GRAAL. The obtained results show that, indeed, AligNet produces biologically more meaningful alignments than state-of-the-art methods and tools, by achieving a better balance between structural matching and protein function conservation. 

We have used both the edge correctness (EC) and the functional coherence (FC) metrics. The results obtained and presented in the Results and Discussion section of this paper, reveal that HubAlign and L-GRAAL obtained the best EC scores when the source input network has considerably less number of edges than the target network, preserving $80\%$ of the edges, while AligNet preserved $60\%$ of the edges.  However, all aligners obtained very low EC scores, with less than $10\%$ of the edges preserved, when the input PPIN have similar size. 

Concerning the efficiency of the considered aligners from the computational point of view, HubAlign and AligNet obtained the best running time. In addition, running time increases with both aligners with the increase in the size of the input networks, unlike the other aligners, which are slower than HubAlign and AligNet and have a variable running time that is not related to the size of the input networks.

%
%
%
%\section*{Funding}
%
%This work was supported by Spanish Ministry of Economy and Competitiveness and European Regional Development Fund project DPI2015-67082-P (MINECO/FEDER) and ``Programa Pont La Caixa per a grups de recerca de la UIB''.
%
%%The research reported in this paper has been partially supported by the Spanish Government, through project DPI2015-67082-P, and the “Programa Pont La Caixa per a grups de recerca de la UIB”.
%
%\section*{Authors' contributions}
%
%ML, FR and GV conceived and coordinated the study, performed data analysis and drafted the manuscript. RA and AA performed all the bioinformatic analyses. All authors read and approved the final manuscript.
%
%\section*{Authors' information}
%
%RA: \texttt{r.alberich@uib.es},
%AA: \texttt{adria.alcala@uib.es},
%ML: \texttt{merce.llabres@uib.es},
%FR: \texttt{cesc.rossello@uib.es},
%GV: \texttt{valiente@cs.upc.edu}.
%
\section*{Acknowledgements}

%We are grateful to the anonymous reviewers for their comments on the paper. 
We thank Gabriel Riera for  the technical support.

%%  After submission of the .TEX file,                     %%
%%  you will be prompted to submit your .BBL file.         %%

% if your bibliography is in bibtex format, use those commands:

\end{document}